\documentclass[conference]{IEEEtran}
\usepackage{graphicx}
\usepackage{wrapfig}
\usepackage{balance}
\usepackage{epsfig}
\usepackage{epstopdf}
\usepackage{indentfirst}
\usepackage{amsmath}
\usepackage{flushend}
\usepackage{makecell}
\usepackage{algorithmic}
\usepackage[linesnumbered,ruled,vlined]{algorithm2e}
\usepackage{subfigure}
\usepackage{hyperref}
\usepackage[normalem]{ulem}
\usepackage{times}
\usepackage[british]{babel}
\usepackage{cite}

\SetKwRepeat{Do}{do}{while}
\newtheorem{definition}{Definition}

\newtheorem{example}{Example}
\newtheorem{lemma}{Lemma}

\newcommand{\State}[1]{#1\\}
\newcommand{\stitle}[1]{\vspace{1ex} \noindent{\bf #1}}

\begin{document}


\title{Finding Top-k Optimal Sequenced Routes\\---Full Version}

\author{\IEEEauthorblockN{Huiping Liu\IEEEauthorrefmark{1},
Cheqing Jin\IEEEauthorrefmark{1},
Bin Yang\IEEEauthorrefmark{2},
Aoying Zhou\IEEEauthorrefmark{1}\\
hpliu@stu.ecnu.edu.cn \quad \{cqjin, ayzhou\}@dase.ecnu.edu.cn \quad byang@cs.aau.dk}
\IEEEauthorblockA{\IEEEauthorrefmark{1}School of Data Science and Engineering, East China Normal University, China}
\IEEEauthorblockA{\IEEEauthorrefmark{2}Department of Computer Science, Aalborg University, Denmark}}



\maketitle

\begin{abstract}
Motivated by many practical applications in logistics and mobility-as-a-service, we study the top-$k$ optimal sequenced routes (KOSR) querying on large, \emph{general} graphs where the edge weights may not satisfy the triangle inequality, e.g., road network graphs with travel times as edge weights. The KOSR querying strives to find the top-$k$ optimal routes (i.e., with the top-$k$ minimal total costs) from a given source to a given destination, which must visit a number of vertices with specific vertex categories (e.g., gas stations, restaurants, and shopping malls) in a particular order (e.g., visiting gas stations before restaurants and then shopping malls).

To efficiently find the top-$k$ optimal sequenced routes, we propose two algorithms \emph{PruningKOSR} and \emph{StarKOSR}. In \emph{PruningKOSR}, we define a dominance relationship between two partially-explored routes. The partially-explored routes that can be dominated by other partially-explored routes are postponed being extended, which leads to a smaller searching space and thus improves efficiency. In \emph{StarKOSR}, we further improve the efficiency by extending routes in an A$^*$ manner. With the help of a judiciously designed heuristic estimation that works for general graphs, the cost of partially explored routes to the destination can be estimated such that the qualified complete routes can be found early. In addition, we demonstrate the high extensibility of the proposed algorithms by incorporating Hop Labeling, an effective label indexing technique for shortest path queries, to further improve efficiency. Extensive experiments on multiple real-world graphs demonstrate that the proposed methods significantly outperform the baseline method. Furthermore, when $k=1$, \emph{StarKOSR} also outperforms the state-of-the-art method for the optimal sequenced route queries.

This is a full version of ``Finding Top-k Optimal Sequenced Routes''~\cite{icde20182}, to appear in IEEE ICDE 2018.
\end{abstract}

\section{Introduction} \label{sec:intro}
Optimal sequenced route (OSR) querying~\cite{osr,osr1}, a.k.a., generalized shortest path querying~\cite{gsp}, aims at finding a route with minimum total cost (e.g., travel distance or travel time), passing through a number of vertex categories (e.g., restaurants, banks, gas stations) in a particular order (e.g., visiting banks before restaurants).
This problem has many practical applications in route planing \cite{mo,icde2018}, crisis management, supply chain management, video surveillance, mobility-as-a-service \cite{trans}, and logistics~\cite{osr,gsp}. 
However, it is often the case that the optimal sequenced route with the minimum total cost may not be the best choice for all users since different users may have different personal preferences \cite{div,Personalized,DBLP:journals/vldb/0002GMJ15}.

\begin{figure}[t]
\centering
\includegraphics[width=2.8in]{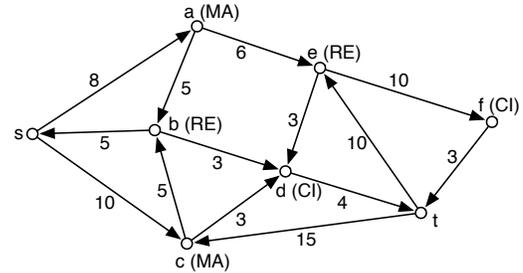}
\caption{A road network graph $G$}
\label{kssr-motivation}
\vspace*{-0.25in}
\end{figure}

Consider the example shown in Figure \ref{kssr-motivation}, where a vertex represents a point-of-interest and is associated with a category, e.g., shopping mall ($\mathit{MA}$), restaurant ($\mathit{RE}$), or cinema ($\mathit{CI}$) and edge weights represent travel costs, e.g., travel time or fuel consumption.
Suppose that Alice plans a trip which starts from location $s$ and wishes passing through a shopping mall, a restaurant, and then a cinema and finally reaching destination $t$. This plan can be formalized with an OSR query with category sequence $\langle \mathit{MA}, \mathit{RE}, \mathit{CI}\rangle$.
The optimal sequenced route for Alice is $s\rightarrow a \rightarrow b\rightarrow d\rightarrow t$ with a cost of 20. However, if Alice prefers restaurant $e$ to restaurant $b$, route $s\rightarrow a \rightarrow e\rightarrow d\rightarrow t$ with a cost of 21 is more preferable. In addition, if the shopping mall at vertex $c$ has sale promotions, route $s\rightarrow c \rightarrow b\rightarrow d\rightarrow t$ with a cost of 22 can also be a good candidate. 
In these cases, returning only the optimal sequenced route may not sufficiently satisfy users' varying preferences. This motivates us to study the top-$k$ optimal sequenced routes (KOSR) querying that returns $k$ routes that satisfy the given category order and have the $k$ least total costs.

In this paper, we focus on finding the top-$k$ optimal sequenced routes in general graphs, where edge weights may not satisfy triangle inequality.
Unfortunately, the KOSR problem on general graphs has not been addressed carefully before, though the OSR problem has been extensively studied. 
In \cite{osr}, the progressive neighbor exploration algorithm {\em PNE} is proposed to solve the OSR problem on general graphs. 
In \cite{gsp}, a dynamic programming based algorithm {\em GSP} is formulated, which outperforms {\em PNE} significantly and is considered as the state-of-the-art for solving the OSR problem on general graphs.

However, by simply extending existing solutions for the OSR problem, it is unlikely, if it is not impossible, to achieve efficient solutions for the KOSR problem.
In particular, dynamic programming based {\em GSP} is unable to be extended to solve the KOSR problem due to lack of sufficient information for other sequenced routes.
%
Although \emph{PNE} can be extended to handle the KOSR problem by iteratively finding the next optimal sequenced route, the efficiency is low since all partially explored sequenced routes whose costs are less than the cost of the $k$-th optimal sequenced route must be examined, whereas most of them can be avoided being extended. 

It is non-trivial to devise an efficient solution for solving KOSR due to two challenges. The first is how to filter unnecessary partially explored sequenced routes when exploring the graph.
To conquer this challenge, we propose a dominance relationship between two partially explored sequenced routes $r$ and $r'$. If $r$ dominates $r'$, the optimal (i.e., least-cost) feasible sequenced route extended from $r$ is always better than that of $r'$. Thus, the exploring of routes that are extended from $r'$ can be postponed until a complete sequenced route extended from $r$ occurs in the result set. 
Furthermore, inspired by A$^*$ algorithm \cite{astar}, we estimate the cost of each partially explored sequenced route to the destination, and explore the partially explored routes according to their estimated total costs, which further reduces the searching space.

The second challenge is how to efficiently find the $i$-th nearest, not merely the nearest, neighbor in a category, as this operation is invoked frequently when solving KOSR. For example, recall that we may want to recommend the top-3 optimal sequenced routes to Alice in Figure \ref{kssr-motivation}.
More than one nearest neighbors in category $\mathit{MA}$ for vertex $s$, i.e., $a$ and $c$, are required to be explored.
%
A simple and intuitive implementation of the operation is to apply Dijkstra's algorithm, which is however very costly. To overcome this weakness, we build an inverted label index for each category by employing hop labeling technique \cite{2hub,hhl,landmark-labeling,hlopt,rpsurvey} on the original graph in an off-line manner. In this way, the $i$-th nearest neighbor in a category can be identified efficiently in an on-line manner by simply looking up the inverted label index.

To the best of our knowledge, this is the first comprehensive work to study the KOSR problem. The paper makes four contributions. First, we propose a dominance relationship between partially explored sequenced routes and develop an algorithm based on the dominance relationship to reduce the searching space significantly when solving the KOSR problem. Second, we propose a heuristic method that is able to estimate the minimal total cost of partially explored sequenced routes, which enables the develop of an A$^*$ like algorithm to further reduce the searching space for solving the KOSR problem. Third, we propose an inverted label index which facilitates the operation that identifies the $i$-th nearest neighbor in a category for a given vertex, which improves the efficiency of both algorithms. Finally, we report on a comprehensive empirical study over different real-world graphs, showing that the proposed algorithms significantly outperform the baseline method for KOSR and the state-of-the-art method for OSR.

\section{Related work} \label{sec:rw}

We categorize relevant studies on sequenced route querying in Table \ref{rw}.
This categorization considers three different aspects.
First, we consider whether the algorithms work for general graphs. When edge weights represent Euclidean distances between vertices, the edge weights satisfy the triangle inequality. We call such graphs \emph{Euclidean graphs}. When edge weights represent other costs such as travel times and fuel consumption~\cite{DBLP:conf/icde/Guo0AJT15, DBLP:conf/icde/YangGJKS14}, the edge weights do not necessarily satisfy triangle inequality anymore. We call such graphs \emph{general graphs}. Note that Euclidean distance and indexing structures based on the Euclidean space, such as R-trees, cannot be utilized in general graphs. The proposed algorithms in this paper work for general graphs.
Second, we consider whether the algorithms support returning the top-$k$ optimal sequenced routes. Most existing studies only work for the case when only the top-1 optimal sequenced route is required. 
Third, we consider whether a specific category order is given.
Table~\ref{rw} clearly shows that this paper is the first comprehensive study for addressing the sequenced route problem on general graphs, with specific category orders, and $k\geq 1$, i.e., the top-$k$ optimal sequenced route (KOSR) problem.


\begin{table}[h]
  \centering
  \small
  \setlength{\tabcolsep}{1pt}
  \vspace*{-0.05in}
  \caption{Categorization of Sequenced Route Queries}
  \begin{tabular} {|c|c|c|} \hline
   & Euclidean Graphs & General Graphs\\ \hline %
  $k=1$ &\makecell[tl]{Specific order:\cite{osr}\\ Arbitrary order:\cite{partialSR,orq,tpq}} & \makecell[tl]{Specific order:\cite{osr,osr1,gsp}\\ Arbitrary order:\cite{tsp,ors}}\\ \hline%
  $k\ge 1$ &\makecell[tl]{Specific order:\cite{gosr,ect,gtp}\\ Arbitrary order:$\emptyset$} & \makecell[tl]{Specific order:~This paper\\ Arbitrary order:$\emptyset$}\\ \hline%
  \end{tabular}
  \label{rw}
\end{table}

The optimal sequenced route querying \cite{osr,osr1}, a.k.a., the generalized shortest path querying \cite{gsp}, is the most relevant problem. 
\cite{osr} is the first work that addresses the problem, in which three algorithms are proposed, namely LORD, R-LORD and PNE. The first two algorithms, LORD and R-LORD, are designed for edge weights in Euclidean spaces where R-trees can be utilized to enable efficient query processing.
The PNE algorithm works for general graphs. In this paper, we extend PNE to solve the KOSR problem, which is regarded as the baseline method.
\cite{osr1} tries to improve the efficiency of optimal sequenced route querying on general graphs by pre-constructing a series of additively weighted voronoi diagrams (AWVD). However, this approach requires a prior knowledge of the category sequence in a query, thus limiting its applicability for online queries, because it is prohibitive to pre-construct AWVDs for all possible category sequences.
%
%
\cite{gsp} addresses the optimal sequenced route queries on general graphs by using a dynamic programming formulation. In their formulation, the optimal costs of all vertices in each category from the start and passing through all the categories before them are computed by using a transition function between consecutive categories. In their solutions, contraction hierarchy technique \cite{ch} is utilized to compute the optimal costs of the vertices in the next category according to above recurrence. Though efficient, this approach cannot be extended to KOSR queries, because the transition function only suits the optimal cost. 


Group optimal sequenced routes problem \cite{gosr,ect,gtp} is also relevant to KOSR. Given a group of users with different sources and destinations and a set of ordered categories, group optimal sequenced routes querying aims to find the top-$k$ optimal sequenced routes that pass through the categories in order and minimize the aggregate travel costs of the group. Specifically, when the group only has one user, then the problem becomes the KOSR problem. However, all existing methods are based on Euclidean space. Thus, they cannot be applied in general graphs. 

\cite{partialSR,orq,tpq,ors} study the problem on finding the optimal route that visits a given set of categories, but without a specific category order. Sometimes, additional constraints, such as partial order 
\cite{orq,partialSR} and budget limit \cite{ors}, are also considered. 
Such problems are NP-hard and can be reduced to generalized traveling salesman problem \cite{tsp}. Therefore, approximate methods are proposed to solve such problems. Due to different problem natures, above methods cannot be directly applied for KOSR. Other advanced routing strategies~\cite{DBLP:journals/sigmod/GuoJ014,DBLP:conf/dasfaa/YangMQZ09,DBLP:journals/tkde/ShangZJYKLW15}, e.g., skyline routing~\cite{DBLP:conf/icde/Guo0AJT15, DBLP:conf/icde/YangGJKS14,DBLP:journals/geoinformatica/Guo0AJT15}, stochastic routing~\cite{pvldb17pathcost, PACE, RiskAware, DBLP:journals/geoinformatica/HuYJM17}, and personalized routing~\cite{DBLP:journals/vldb/0002GMJ15,Personalized}, are also different from KOSR.

\section{Preliminaries} \label{sec:pro}

We formalize the KOSR problem and introduce baseline solution. Frequent notations are summarized in Table \ref{notations}.

\begin{table}[!t]
  \centering
  \scriptsize
  \setlength{\tabcolsep}{1pt}
  \vspace*{-0.2cm}
  \caption{Notation}
  \begin{tabular} {|c|c|} \hline
  \textbf{Notation} & \textbf{Meaning}\\ \hline %
  $P_{s,t}$ & A route from $s$ to $t$ \\  \hline
  $C$ & A category sequence $C= \langle C_1,\cdots$, $C_j\rangle$\\  \hline
  $|C|$ & The number of categories in category sequence $C$ \\  \hline
  $V_{C_i}$ & The vertex set of category $C_i$ \\ \hline
  $|C_i|$ & The number of vertices that belong to category $C_i$, i.e., $|V_{C_i}|$  \\  \hline
  $P_{s,t,C}$ & \makecell[tl]{Witness $P_{s,t,C=\langle C_1,\cdots,C_j\rangle}=\langle s,v_1,\cdots,v_j,t\rangle$, such that \\$v_i\in V_{C_i}$ for $1\le i\le j$} \\ \hline
  $|P|$ & The number of vertices in route or witness $P$ \\ \hline
  $w(P)$ & The weight of route or witness $P$ \\ \hline
  $dis(v_i,v_j)$ & The least cost from vertex $v_i$ to $v_j$ \\ \hline
  $k$ & Top $k$ results are needed \\ \hline
  \end{tabular}
  \label{notations}
  \vspace*{-0.4cm}
\end{table}

\subsection{Problem Definition}

\begin{sloppypar}
\begin{definition}
[Graph] A directed weighted graph $G(V, E, F, W)$ includes a vertex set $V$ and an edge set $E\subseteq V\times V$. Category function $F: V\rightarrow 2^{S}$ takes as input a vertex $v\in V$ and returns a set of categories $F(v)$, where $S$ denotes a set of all possible categories. Weight function $W: E\rightarrow {R}^+$ takes as input an edge $(u, v)$ and returns a non-negative cost of the edge $W((u,v))$, e.g., the travel time when traversing edge $(u, v)$.
%
\end{definition}
\end{sloppypar}

For example, in Figure~\ref{kssr-motivation}, we have $S=\{\mathit{MA}, \mathit{RE}, \mathit{CI}\}$, $F(a)=\{\mathit{MA}\}$, and $W((s, a))=8$. Note that the edge weights can be arbitrary and may not satisfy the triangle inequality. 

\begin{definition}
[Route] A route $P_{s,t}$ from vertex $s$ to vertex $t$ in graph $G$ is a sequence of vertices, where each two adjacent vertices are connected by an edge, denoted by $P_{s,t}=\langle v_0=s, v_1,\cdots, v_q=t\rangle$. Let $w(P_{s,t})=\sum_{0\le i< q}{W((v_i,v_{i+1}))}$ be the weight, or cost, of route $P_{s,t}$ and $|P_{s,t}|$ be the size of route $P_{s,t}$ which equals to the number of vertices in route $P_{s,t}$.
\end{definition}

\begin{definition}
[Category Sequence] A category sequence $C= \langle C_1, C_2,\cdots$, $C_j\rangle$ represents an order in which each category must be visited, where each $C_i\in C$, $1\le i\le j$, represents a specific category in category set $S$, and each $C_i$ corresponds to a vertex set $V_{C_i}=\{v|v\in V \bigwedge C_i\in F(v)\}$. We refer $|C|$ and $|C_i|$ to the size of the category sequence $C$ and the size of $V_{C_i}$, respectively. 
\end{definition}

\footnotetext[1]{Note that a witness may not represent a route according to Definition 2 as consecutive vertices in a witness may not be connected by an edge.}

\begin{definition}
[Feasible Route] Given a source-destination pair $(s, t)$, and a category sequence $C=\langle C_1, C_2,\cdots, C_j\rangle$, a route $P_{s,t}= \langle v_0=s, v_1,\cdots, v_q=t\rangle$ is feasible if and only if there exists a subsequence of vertices $\langle v_{r_1}, v_{r_2},\cdots, v_{r_j}\rangle$ from $P_{s,t}$, such that $0< r_1\le r_2\le \cdots \le r_j< q$ and for $1\le i\le j$, $v_{r_i}\in V_{C_i}$ or $C_i\in F(v_{r_i})$. We call $\langle s,v_{r_1}, v_{r_2},\cdots, v_{r_j},t\rangle$ the witness\footnotemark[1] of $P_{s,t}$ w.r.t category sequence $C$, denoted as $P_{s,t,C}$.
\end{definition}

In many cases, there exist multiple feasible routes for a given source-destination pair and a category sequence. We distinguish two feasible routes according to their witnesses. This means that if two feasible routes share the same witness w.r.t a category sequence, they are regarded as the same feasible route and only the route with smaller cost is considered. Formally, for a witness $P_{s,t,C}=\langle v_0, v_1,\cdots, v_q\rangle$, its cost $w(P_{s,t,C})$ is defined as $w(P_{s,t,C})=\sum_{0\le i< q}{dis(v_i,v_{i+1})}$, where $dis(v_i,v_{i+1})$ is the least cost from vertex $v_i$ to $v_{i+1}$.


\begin{definition}
[KOSR query] Given a graph $G$, the top-$k$ optimal sequenced routes (KOSR) query is a quad-tuple $(s, t, C, k)$, where $s,t\in V$ denotes a source-destination pair, $C$ is a category sequence, and $k$ is a positive integer.
The query returns a set of $k$ different feasible routes w.r.t $C$, $\Psi=\{P^1_{s,t}, P^2_{s,t},\cdots, P^k_{s,t}\}$, such that there does not exist any other feasible route $P'_{s,t}$ in $G$ where $P'_{s,t}\notin \Psi \bigwedge w(P'_{s,t})< \max_{1 \leq i \leq k}{w(P_{s,t}^i)}$. 
\end{definition}

\begin{sloppypar}
\begin{example}
Consider the graph $G$ in Figure \ref{kssr-motivation}, the KOSR query $(s,t, \langle \mathit{MA},\mathit{RE},\mathit{CI}\rangle, 3)$ returns $\Psi$=$\{\langle s,a,b,d,t\rangle$, $\langle s,a,e,d,t\rangle$, $\langle s,c,b,d,t\rangle\}$ that includes routes with costs of 20, 21, and 22. There does not exist another feasible path whose cost is smaller than 22.
\end{example}
\end{sloppypar}

To simplify later discussion, we focus on identifying the witnesses of top-$k$ optimal sequenced routes, rather than identifying the actual routes. However, given the witness, its actual route can be easily reconstructed.  
For simplicity, all routes we discuss in the following sections refer to witnesses unless stated otherwise.
Moreover, given a category sequence $C$, we introduce two dummy categories $C_0=\{s\}$ and $C_{|C|+1}=\{t\}$ to include the source vertex $s$ and destination vertex $t$. 


\subsection{Baseline Solution} \label{sub:pne}
Since OSR can be considered as a special case of KOSR where $k$ is set to 1, we present PNE and GSP, two state-of-the-art methods for solving OSR, and we present the baseline KPNE, which is extended from PNE, for solving KOSR.

\subsubsection{PNE} \label{sub:gsp}
\begin{sloppypar}
The progressive neighbor exploration (PNE) algorithm \cite{osr} is able to find the optimal sequenced route in general graphs. Algorithm \ref{pne} shows the sketch of PNE. During the processing, a priority queue is maintained for partially explored routes (witnesses). At each iteration, the route $\langle v_0,v_1,\cdots,v_{q-1},v_q\rangle$ with minimal cost in the priority queue is chosen to be examined, where $v_i \in V_{C_i}$ for each $1\le i\le q$. To extend from the route, we need to consider vertices in the next category $C_{q+1}$. Instead of extending the route via all its neighbors in category $C_{q+1}$, only the nearest neighbor $v_{q+1}$ of $v_q$, such that $dis(v_q,v_{q+1})=\arg \min_{v\in V_{C_{q+1}}} dis(v_q, v)$, is considered. Moreover, to guarantee the correctness, another candidate route derived from $\langle v_0,v_1,\cdots,v_{q-1},v_q\rangle$ is incrementally generated by extending $\langle v_0,v_1,\cdots,v_{q-1}\rangle$ via $v_{q-1}$'s next nearest neighbor $v'_q$ in $C_q$, such that $v'_q\ne v_q$ and $dis(v_{q-1},v'_q)\ge dis(v_{q-1},v_q)$. The algorithm returns the optimal route as it passes through all categories in order and reaches the destination. Since a vertex's neighbors in the next category $C_i (1\le i\le j)$ can be as many as $|C_i|$, it is impractical to compute the least costs from the vertex to all its neighbors. By progressively extending route via its nearest neighbors and generating candidate route derived from it, PNE carefully examines all the possible partially explored candidate routes on demand to find the optimal sequenced route. It is possible to extend PNE to solve KOSR problem, we only need to add a result set and each time we find an optimal sequenced route (line 5), it will be added to the result set, when the result set consists of $k$ routes or the priority queue is already empty, the set will be returned as the result of KOSR. We refer to this method for solving KOSR as \emph{KPNE}.
\end{sloppypar}

\begin{algorithm}[t]
\caption{$PNE(G,s,t,C)$ 
}
\label{pne}

\KwIn {
 Graph $G(V,E)$, source-destination pair $s,t\in V$, category sequence $C= \langle C_1,\cdots$, $C_j\rangle$.}

\KwOut {
The optimal sequenced route.}

\State {Priority queue $Q \leftarrow \{\langle s\rangle\}$;}

\While {$|Q|>0$}{
    \State {$\langle v_0=s,v_1,\cdots,v_{q-1},v_q\rangle \leftarrow Q.\textsf{extractMin}()$;}
    \If {$q=|C|+1$}{
        \Return {$\langle v_0,v_1,\cdots,v_{q-1},v_q\rangle$;}
    }
    \tcp {extend route.}
    \State {$v_{q+1}\leftarrow v_q$'s nearest neighbor in category $C_{q+1}$;}
    \State {$Q.\textsf{insert}(\langle v_0,v_1,\cdots,v_q,v_{q+1}\rangle)$;}
    \tcp {generate candidate route.}
    \If {$q>0$}{
        \State {$v'_q \leftarrow v_{q-1}$'s next nearest neighbor in $C_{q}$;}
        \tcp{$v'_q\ne v_q \bigwedge dis(v_{q-1},v'_q)\ge dis(v_{q-1},v_q)$}
        \State {$Q.\textsf{insert}(\langle v_0,v_1,\cdots,v_{q-1},v'_q\rangle)$;}
    }
}
\end{algorithm}

\subsubsection{GSP} \label{sub:gsp}
\begin{sloppypar}
 Another state-of-the-art method, namely \emph{GSP}, for the optimal sequenced route within graph is proposed in \cite{gsp}, in which, a dynamic programming solution is formulated as follows:
$$X[i,j]=
\begin{cases}
0  &\text{$i=0$}\\
\min \limits_{1\le l\le |C_{i-1}|}\{X[i-1,l]+dis(v_{i-1,l},v_{i,j})\} &\text{$i>0$},
\end{cases}
$$
where $X[i,j]$  records the least cost of the route of the $j$-th vertex (starting from 0) in the $i$-th category that from the source and passes through all the categories before it and $dis(v_{i-1,l},v_{i,j})$ is the least cost from the $l$th vertex in category $C_{i-1}$ to the $j$-th vertex in category $C_i$. As a result, the cost of the optimal sequenced route will be $X[|C|+1,0]$. To compute the matrix $X$ efficiently, contraction hierarchy technique \cite{ch} is utilized to compute the least costs of the vertices in the next category according to above recurrence. By applying $O(|C|)$ times of forward search (Dijkstra's algorithm based search by using contraction hierarchy) and backward search (DFS based) as well as pruning optimizations, GSP can efficiently find the optimal sequenced route. However, since only the least cost of each vertex is considered and the above recurrence only suits the least cost, GSP cannot be directly extended to KOSR problem.
\end{sloppypar}

\begin{sloppypar}
Although KPNE which is extended from \cite{osr} is able to solve KOSR on general graphs, it is inefficient since all partially explored candidate routes whose costs are smaller than the cost of the $k$-th optimal sequenced route must be examined. In the worst case, the number of examined partially explored candidate routes at category $C_i$ can reach $\prod_{1\le j\le i}|C_j|$, as a result, the total number of routes to be examined by KPNE can be $\sum_{1\le i\le |C|+1}\prod_{1\le j\le i}|C_j|$, which is too huge to process on large graphs.
\end{sloppypar}

\section{Proposed solutions for KOSR} \label{sec:sol}


In this section, we propose two efficient methods to solve KOSR. We first describe a method based on the route \emph{dominance} relationship to filter unnecessary partially explored candidate routes in Section \ref{prunmethod}, which reduces the searching space. Moreover, we demonstrate the extensibility of the proposed method by incorporating an optimization technique that is able to find the $i$-th nearest neighbor in a category for a given vertex efficiently. Subsequently, we further reduce the searching space by integrating a heuristic estimation in an A$^*$ manner in Section \ref{targetmethod}. 

\subsection{Dominance Based Algorithm} \label{prunmethod}

We first illustrate the intuition of the route dominance relationship. Consider a KOSR query $(s,t,\langle \mathit{MA},\mathit{RE},\mathit{CI}\rangle,2)$ in Figure \ref{kssr-motivation}. In order to find the first optimal sequenced route $\langle s,a,b,d,t\rangle$ with the cost of 20 (shorten as $\langle s,a,b,d,t\rangle (20)$), KPNE will attempt to examine and extend $\langle s,a,b\rangle (13)$ and $\langle s,c,b\rangle (15)$, because both $\langle s,a,b\rangle$ and $\langle s,c,b\rangle$ have a smaller cost than $\langle s,a,b,d,t\rangle$. However, there is no need to extend $\langle s,c,b\rangle$ to find $\langle s,a,b,d,t\rangle$, because the cost of the optimal feasible route extended from $\langle s,c,b\rangle$ won't be smaller than that of $\langle s,a,b\rangle$ (i.e., $\langle s,a,b,d,t\rangle$). Hence, $\langle s,c,b\rangle$ can be excluded to be extended until the optimal sequenced route $\langle s,a,b,d,t\rangle$ is found. In this case, we say $\langle s,c,b\rangle$ is \emph{dominated} by $\langle s,a,b\rangle$. Next, we formally define the dominance relationship.

\begin{sloppypar}
\begin{definition}
[Dominance] Consider a given category sequence $C=\langle C_1,\cdots,C_j\rangle$ and two partially explored candidate routes (witnesses) $P_1=\langle s,v^1_1,\cdots,v^1_q\rangle$ and $P_2=\langle s,v^2_1,\cdots,v^2_q\rangle$ ($1\le q\le j$). If $v^1_q=v^2_q$ and $w(P_1)\le w(P_2)$ holds, $P_1$ dominates $P_2$ w.r.t $C$, denoted as $P_1\prec_C P_2$.
\end{definition}
\end{sloppypar}


\begin{lemma} \label{prune}
Given a KOSR query $(s,t,C$=$\langle C_1,\cdots,C_j\rangle,k)$ and two partially explored routes $P_1$ and $P_2$, if $P_1 \prec_C P_2$, then $w(P^*_1)\le w(P^*_2)$, where $P^*_1$ and $P^*_2$ are the optimal feasible routes that are extended from $P_1$ and $P_2$, respectively.
\end{lemma}

\begin{proof}
Suppose $P_1$=$\langle s,v^1_1,\cdots,v^1_q\rangle$, $P_2$=$\langle s,v^2_1,\cdots,v^2_q\rangle$ $(1\le q\le j)$ and $P^*_1$=$\langle s,v^1_1,\cdots,v^1_q,v_{q+1},\cdots,v_j, t\rangle$, since $P^*_1$ is the optimal feasible route extended from $P_1$, $P=\langle v^1_q,v_{q+1},\cdots,v_j,t\rangle$ must be the optimal sequenced route for category sub-sequence $\langle C_{q+1}, \cdots, C_j \rangle$ from $v^1_q$ to $t$. Because $P_1 \prec_C P_2$, we have $v^2_q=v^1_q$ and $w(P_1)\le w(P_2)$, thus, $P^*_2$ can be represented by $\langle s,v^2_1,\cdots,v^2_q,v_{q+1},\cdots,v_j,t\rangle$, then $w(P^*_1)=w(P_1)+w(P)$ and $w(P^*_2)=w(P_2)+w(P)$, since $w(P_1)\le w(P_2)$, we have $w(P^*_1)\le w(P^*_2)$. %
\end{proof}

According to Lemma \ref{prune}, there is no need to extend the dominated partially explored routes until the optimal feasible route extended from their dominating route become one of the top-$k$ optimal sequenced routes. This is because the partially explored candidate routes that are dominated by other partially explored candidate routes with smaller costs can never be extended to be the next optimal sequenced routes before their dominating routes. On the other hand, after an optimal sequenced route is found, we need to reconsider its corresponding dominated routes, so that they can be extended to be the next optimal sequenced routes. Based on the dominance relationship, we propose PruningKOSR method (Algorithm \ref{pruningkssr}).

\begin{algorithm}[t]
\caption{$PruningKOSR(G,s,t,C,k)$ 
}
\label{pruningkssr}

\KwIn {
 Graph $G(V,E)$, source-destination pair $s,t$$\in V$, category sequence $C= \langle C_1, \cdots$, $C_j\rangle$, and $k$.}

\KwOut {
The top-$k$ optimal sequenced routes.}

\State {$\forall v\in V$, initialize $v.\textbf{HT}_{\prec_C}$ and $v.\textbf{HT}_{\succ_C}$;}
\State {$\Psi \leftarrow \emptyset$;}
\State {Priority queue $Q \leftarrow \{(\langle s\rangle, 1)\}$;\tcc*[f] {($route,x$)}}

\While {$Q$ is not empty and $|\Psi|<k$ }{
    \State {$p=(\langle v_0,v_1,\cdots,v_{q-1},v_q\rangle,x) \leftarrow Q.\textsf{extractMin}()$;}
    \If {$q=|C|+1$}{
        \State {$\Psi \leftarrow \Psi \bigcup \{p\}$;}
        \tcp {reconsider dominated routes}

        \For {each $i=1\cdots q-1$}{
            \If {$\langle v_0,\cdots,v_i\rangle = v_i.\textbf{HT}_{\prec_C}.\textsf{getValue}(i+1)$}{
                \State {$p'=(\langle v_0,v'_1,\cdots,v_i\rangle,x)\leftarrow v_i.\textbf{HT}_{\succ_C}.\textsf{getValue}(i+1).\textsf{extractMin}()$;}
                \State {$Q.\textsf{insert}(p'=(\langle v_0,v'_1,\cdots,v_i\rangle,-))$;}
                \State {$v_i.\textbf{HT}_{\prec_C}.\textsf{remove}(i+1)$;}
            }
        }
    }
    \Else {
        \tcp {pruning dominated routes}
        \If {$|p|\notin v_q.\textbf{HT}_{\prec_C}.\textsf{KeySet}$}{
            \State {$v_q.\textbf{HT}_{\prec_C}.\textsf{add}(|p|,\langle v_0,\cdots,v_q\rangle)$;}
            \State {$v_{q+1}\leftarrow \textsf{FindNN}(v_q,C_{q+1},1)$;}
            \State {$Q.\textsf{insert}((\langle v_0,v_1,\cdots,v_q,v_{q+1}\rangle, 1))$;}
        }
        \Else {$v_q.\textbf{HT}_{\succ_C}.\textsf{getValue}(|p|).\textsf{insert}(p)$;}
        \If {$q>0$}{
            \State {$v'_q \leftarrow \textsf{FindNN}(v_{q-1},C_q,x+1)$;}
            \State {$Q.\textsf{insert}((\langle v_0,v_1,\cdots,v_{q-1},v'_q\rangle,x+1))$;}
        }
    }
}
\Return {$\Psi$;}
\end{algorithm}

To check the dominance relationship and maintain the dominated routes, for each vertex $v$, we introduce two hash tables in the form of $(key, value)$ pairs. One is \textbf{HT}$_{\prec_C}$ for dominating routes, where $key$ is the size of the partially explored dominating route that has been extended at $v$, and the $value$ is the route itself. Another one is \textbf{HT}$_{\succ_C}$ for dominated routes, where $key$ represents the size of dominated route, and $value$ is a priority queue for the routes with the size of $key$ that have reached $v$ and been dominated, the dominated routes are ordered according to their costs in an ascending order. We also maintain a result set $\Psi$ for the top-$k$ optimal sequenced routes and a global priority queue $Q$ for partially explored routes (witnesses) sorted by their costs in an ascending order. Moreover, for each route $p=\langle v_0,v_1,\cdots,v_{q-1},v_q\rangle$, we introduce an additional attribute $x$ to indicate that $v_q$ is the $x$-th nearest neighbor of $v_{q-1}$ in category $C_q$ when generating $p$.  Initially, only the source with $x=1$ is added to the queue $Q$. Then we begin a loop until $Q$ is empty or the top-$k$ optimal sequenced routes have been found.

\stitle {Pruning dominated routes:} At each iteration, the route with the minimum cost is chosen to be examined. If it already reaches the destination, we add it to the result set and reconsider the dominated routes (lines 6--12). Otherwise, we check whether it is dominated. For a route $p=\langle v_0,v_1,\cdots,v_{q-1},v_q\rangle$ to be examined, if $p$ is the first route with size $|p|$ that reaches vertex $v_q$, we add $p$ to the \textbf{HT}$_{\prec_C}$ of $v_q$ and extend it via $v_q$'s nearest neighbor in category $C_{q+1}$ (lines 14--17). Otherwise, if its size $|p|$ is in the \textbf{HT}$_{\prec_C}$ of $v_q$, it means that another route with size $|p|$ and smaller cost has been reached and extended at $v_q$, so that $p$ is dominated. According to Lemma \ref{prune}, there is no need to extend $p$ anymore, therefore, we insert it into the \textbf{HT}$_{\succ_C}$ of $v_q$ instead of the priority queue $Q$(line 19). Subsequently, we generate a new candidate route derived from $p$. Since the candidate route via the $x$-th nearest neighbor of $v_{q-1}$ has been generated in previous iterations, we need to find $v_{q-1}$'s $(x+1)$-th nearest neighbor in category $C_{q}$, $v'_q$, by invoking algorithm \textsf{FindNN}, and create candidate route $\langle v_0,v_1,\cdots,v_{q-1},v'_q\rangle$ with incremental $x$ and insert it into the priority queue (lines 20--22).

\stitle {Reconsider dominated routes:} After an optimal sequenced route $p$ has been found, we need to reconsider the partially explored routes that are dominated by sub-routes of $p$, since these routes now can possibly be extended to be the next optimal sequenced route. Therefore, for each vertex $v_i$ in $p$, if $\langle v_0,\cdots,v_i\rangle$ dominates the routes with size of $i+1$ in the \textbf{HT}$_{\succ_C}$ of $v_i$ (line 9), we only reconsider the dominated route $p'$ with the least cost in the \textbf{HT}$_{\succ_C}$ of $v_i$, because other routes in \textbf{HT}$_{\succ_C}$ of $v_i$ are dominated by $p'$. This also explains why we use a priority queue as the $value$ in hash table \textbf{HT}$_{\succ_C}$. Since $p'$'s $x+1$ nearest neighbor has been computed after it is dominated, we set its $x$ to `-' (which means there is no need to generate candidate route that is derived from $p'$) and re-add it to the priority queue (lines 10--11). Meanwhile, we remove $\langle v_0,\cdots,v_i\rangle$ from the \textbf{HT}$_{\prec_C}$ of $v_i$, so that the next candidate route that reaches $v_i$ can be extended (line 12).

\begin{table}[t]
  \centering
  \small
  \vspace*{-0.1in}
  \caption{Running example of Alg. \ref{pruningkssr} for Fig. \ref{kssr-motivation}}
  \subtable[Routes in the priority queue $Q$]{
  \setlength{\tabcolsep}{1pt}
    \begin{tabular}{cc} \hline
    \textbf{Step} & \textbf{Routes (route(cost), x)}\\ \hline
    1 & \makecell[tl]{$(\langle s\rangle(0),1)$}  \\
    2 & \makecell[tl]{$(\langle s,a\rangle(8),1)$} \\
    3 & \makecell[tl]{$(\langle s,c\rangle(10),2)$,$(\langle s,a,b\rangle(13),1)$}\\
    4 & \makecell[tl]{$(\langle s,a,b\rangle(13),1)$,$(\langle s,c,b\rangle(15),1)$}\\
    5 & \makecell[tl]{$(\langle s,a,e\rangle(14),2)$,$(\langle s,c,b\rangle(15),1)$,$(\langle s,a,b,d\rangle(16),1)$}\\
    6 & \makecell[tl]{\uwave{$(\langle s,c,b\rangle(15),1)$},$(\langle s,a,b,d\rangle(16),1)$,$(\langle s,a,e,d\rangle(17),1)$}\\
    7 & \makecell[tl]{$(\langle s,a,b,d\rangle(16),1)$,$(\langle s,a,e,d\rangle(17),1)$,$(\langle s,c,e\rangle(27),2)$}\\
    8 & \makecell[tl]{\uwave{$(\langle s,a,e,d\rangle(17),1)$},$(\langle s,a,b,d,t\rangle(20),1)$,$(\langle s,c,e\rangle(27),2)$,\\$(\langle s,a,b,f\rangle(40),2)$}\\
    9 & \makecell[tl]{\underline{$(\langle s,a,b,d,t\rangle(20),1)$},$(\langle s,a,e,f\rangle(24),2)$,$(\langle s,c,e\rangle(27),2)$,\\$(\langle s,a,b,f\rangle(40),2)$}\\
    10 & \makecell[tl]{$(\langle s,c,b\rangle(15),-)$,$(\langle s,a,e,d\rangle(17),-)$,$(\langle s,a,e,f\rangle(24),2)$,\\$(\langle s,c,e\rangle(27),2)$,$(\langle s,a,b,f\rangle(40),2)$}\\
    11 & \makecell[tl]{$(\langle s,a,e,d\rangle(17),-)$,$(\langle s,c,b,d\rangle(18),1)$,$(\langle s,a,e,f\rangle(24),2)$,\\$(\langle s,c,e\rangle(27),2)$,$(\langle s,a,b,f\rangle(40),2)$}\\
    12 & \makecell[tl]{$(\langle s,c,b,d\rangle(18),1)$,$(\langle s,a,e,d,t\rangle(21),1)$,$(\langle s,a,e,f\rangle(24),2)$,\\$(\langle s,c,e\rangle(27),2)$,$(\langle s,a,b,f\rangle(40),2)$}\\
    13 & \makecell[tl]{\underline{$(\langle s,a,e,d,t\rangle(21),1)$},$(\langle s,c,b,d,t\rangle(22),2)$,$(\langle s,a,e,f\rangle(24),2)$,\\$(\langle s,c,e\rangle(27),2)$,$(\langle s,a,b,f\rangle(40),2)$,$(\langle s,c,b,f\rangle(42),2)$}\\ \hline
    \end{tabular}
    \label{queue_example}
    }
    \subtable[Hash tables of vertex $b$ with respect to Table \ref{queue_example}]{
    \begin{tabular}{c|c|c} \hline
    \textbf{Step} & \textbf{HT}$_{\prec_C}$ & \textbf{HT}$_{\succ_C}$\\ \hline
    1& $\emptyset$ & $\emptyset$ \\
    4& \quad\quad$(3,\langle s,a,b \rangle)$ \quad\quad& $\emptyset$ \\
    6& \quad\quad$(3,\langle s,a,b \rangle)$ \quad\quad& \quad $(3,\{\langle s,c,b \rangle(15)\})$\quad \\
    9& $\emptyset$ & $(3,\{\})$ \\
    10& \quad\quad$(3,\langle s,c,b \rangle)$ \quad\quad& $(3,\{\})$ \\ \hline
    \end{tabular}
    \label{HT_example}
    }
  \label{kssr_example}
  \vspace*{-0.3in}
\end{table}


\begin{lemma}
Algorithm \ref{pruningkssr} returns the correct result for a KOSR query.
\end{lemma}

\begin{proof}
  To find the next optimal sequenced route, all possible partially explored candidate routes are considered (lines 14--17 and 20--22) except for the dominated routes (line 19) which can be removed from extending according to Lemma \ref{prune}. After an optimal sequenced route is found, the dominated routes that can be extended to be the next optimal sequenced route are reconsidered. Therefore, Algorithm \ref{pruningkssr} returns the correct result for a KOSR query.
\end{proof}

\begin{example} \label{prune_example}
Consider Figure \ref{kssr-motivation}. Suppose the given query is $(s,t,\langle \mathit{MA},\mathit{RE},\mathit{CI}\rangle,2)$. Table \ref{kssr_example} shows the routes in the priority queue $Q$ at each step and the hash tables of vertex $b$ at different steps. At step 1, route $\langle s\rangle$ is added to the queue, then it is extended via $a$ ($s$'s nearest neighbor in category $\mathit{MA}$), and no candidate route can be generated. At step 2, $\langle s,a\rangle$ is examined, it is extended via $b$ ($a$'s nearest neighbor in category $\mathit{RE}$) and candidate route $\langle s,c\rangle$ is generated via $s$'s 2nd nearest neighbor in category $\mathit{MA}$. At step 4, $\langle s,a,b\rangle$ is examined and extended at $b$, we insert it into the \textbf{HT}$_{\prec_C}$ of $b$. Subsequently, at step 6, since $\langle s,c,b\rangle$ is dominated by $\langle s,a,b\rangle$ in the \textbf{HT}$_{\prec_C}$ of $b$ , $\langle s,c,b\rangle$ won't be extended at $b$, instead, we insert $\langle s,c,b\rangle$ into the \textbf{HT}$_{\succ_C}$ of $b$, and generate candidate route $\langle s,c,e\rangle$ via $c$'s 2nd nearest neighbor $e$ in category $\mathit{RE}$. At step 9, the first optimal sequenced route $\langle s,a,b,d,t\rangle$ is found. Since both $\langle s,c,b\rangle$ and $\langle s,a,e,d\rangle$ in \textbf{HT}$_{\succ_C}$ of $b$ and $d$, respectively, are dominated by $\langle s,a,b\rangle$ and $\langle s,a,b,d\rangle$ in \textbf{HT}$_{\prec_C}$ of $b$ and $d$, respectively, we re-add them into the queue with $x$=`-' and remove the corresponding dominating routes from \textbf{HT}$_{\prec_C}$. Finally, at step 13, the second optimal sequenced route $\langle s,a,e,d,t\rangle$ is found, and we return $\{\langle s,a,b,d,t\rangle,\langle s,a,e,d,t\rangle\}$ as the result.
\end{example}

By pruning the dominated routes and the candidate routes derived from them, both the capacity of the priority queue and the searching space are reduced, which improves the efficiency. Given a KOSR query $\langle s,t,C,k \rangle$, to find the first optimal sequenced route, for each vertex $v$ in $C_i (0\le i\le |C|)$, at most one route with size $(i+1)$ (plus the source) is extended at $v$ (line 16 in Algorithm \ref{pruningkssr}), and at most $|C_{i+1}|-1$ candidate routes can be generated via $v$'s next nearest neighbors in category $C_{i+1}$ (line 21 in Algorithm \ref{pruningkssr}). As a result, in the worst case, the number of routes to be examined by Algorithm \ref{pruningkssr} for the first optimal sequenced route is $\sum_{0\le i\le |C|}|C_i|\cdot |C_{i+1}|$, in which $\sum_{0\le i\le |C|}|C_i|$ routes are extended. Then, for each of the next $k-1$ optimal sequenced routes, at most $|C|$ dominated routes are reconsidered once an optimal sequenced route is found, which results in at most $\sum_{2\le i\le |C|+1}|C_i|$ examined routes, and in which at most $|C|$ routes are extended at $|C|$ different categories, respectively. That is, to find the top-$k$ optimal sequenced routes, at most $\sum_{0\le i\le |C|}|C_i|\cdot |C_{i+1}|+(k-1)\cdot\sum_{2\le i\le |C|+1}|C_i|$ partially explored routes need to be examined, in which $\sum_{0\le i\le |C|}|C_i|+(k-1)\cdot|C|$ routes are extended. Compared to KPNE, the searching space is reduced from exponential complexity ($\sum_{1\le i\le |C|+1}\prod_{1\le j\le i}|C_j|$) down to polynomial complexity ($\sum_{0\le i\le |C|}|C_i|\cdot |C_{i+1}|+(k-1)\cdot\sum_{2\le i\le |C|+1}|C_i|$). Lemma \ref{complexity} shows the time complexity of Algorithm \ref{pruningkssr}.


\begin{sloppypar}
\begin{lemma} \label{complexity}
Given a KOSR query $(s,t,C,k)$, let $M=\sum_{0\le i\le |C|}|C_i|\cdot |C_{i+1}|+(k-1)\cdot\sum_{2\le i\le |C|+1}|C_i|,N=\sum_{0\le i\le |C|}|C_i|+(k-1)\cdot|C|$, the time complexity of Algorithm \ref{pruningkssr} is $O(M\rho+M\log N)$, where $\rho$ is the time complexity of Algorithm \textsf{FindNN}.
\end{lemma}
\end{sloppypar}

\begin{proof}
Since at most $M$ partially explored candidate routes are generated during the process of Algorithm \ref{pruningkssr}, which means Algorithm \textsf{FindNN} will be called $M$ times at most, in which, at most $N$ routes are extended via the nearest neighbor. So that the complexity of this part is $O(M\rho)$. In addition, each time we examine a candidate route from the priority queue, if the route is extended via the nearest neighbor, two candidate routes are generated in total, in this case, the capacity of the priority queue will be increased by 1. Otherwise, if the route is dominated, then it cannot be extended and only one candidate route is generated via the next nearest neighbor, and the capacity of the priority queue will not change. Since at most $N$ candidate routes are extended via their nearest neighbors, the capacity of the priority queue is at most $N$. As a result, the complexity of the maintenance of the priority queue is $O(M\log N)$. In summary, the total time complexity of Algorithm \ref{pruningkssr} is $O(M\rho+M\log N)$. 
\end{proof}


\stitle {Finding the $x$-th nearest neighbor}. Next, we introduce how to find the $x$-th nearest neighbor, the core operation \textsf{FindNN} in PruningKOSR. A straightforward way to find the $x$-th nearest neighbor of vertex $v_i$ in category $C_{i+1}$ is that by using Dijkstra's search. We start from $v_i$ and extend vertices via their adjacent vertices until the $x$-th vertex in $V_{C_{i+1}}$ is settled. However, each time we find the $x$-th nearest neighbor, Dijkstra's search actually finds the top-$x$ nearest neighbors from scratch, which results in duplicate search effort throughout the graph. Moreover, since \textsf{FindNN} is frequently invoked, frequent Dijkstra's searches on large graphs are practically inefficient. Hence, a more efficient method without duplicate searches is called for. To this end, we propose a method to incorporate the use of 2-hop labeling technique \cite{2hub,hhl,landmark-labeling} to find the $x$-th nearest neighbor.

Given a directed weighted graph $G(V,E)$, for each vertex $v\in V$, 2-hop labeling maintains two labels $L_{in}(v)$ and $L_{out}(v)$. In particular, $L_{in}(v)$ consists of a set of label entries in the form of $(u, d_{u,v})$, where $u\in V$ is a vertex that is able to reach $v$, and $d_{u,v}$=$dis(u, v)$. Similarly, $L_{out}(v)$ consists of a set of label entries in the form of $(u', d_{v, u'})$, where $u'\in V$ is a vertex that can be reached by $v$, and $d_{v, u'}$=$dis(v, u')$. Note that $L_{in}(v)$'s entries may only contain a subset of vertices that can reach $v$; similarly, $L_{out}(v)$'s entries may only contain a subset of vertices that can be reached by $v$.
%
%
In addition, the labels must satisfy the \emph{cover property}: for any two vertices $s$ and $t$, there exists a vertex $u$ on the shortest path from $s$ to $t$ that belongs to both $L_{out}(s)$ and $L_{in}(t)$. Based on which, to answer a least cost query from $s$ to $t$, we compute as follows:
\begin{equation}
\nonumber
dis(s,t)=\min \{d_{s,u}+d_{u,t}|(u,d_{s,u})\in L_{out}(s), (u,d_{u,t})\in L_{in}(t)\}.
\end{equation}
Hence, the least cost from $s$ to $t$ can be computed by scanning $L_{out}(s)$ and $L_{in}(t)$ to find their matching label entries. If the label entries in each label set are sorted by their vertices, then we can compute $dis(s,t)$ in $O(|L_{out}(s)|+|L_{in}(t)|)$ time using a merge-join like algorithm.

We note that building the 2-hop labeling with the minimal size (where the size of the index is defined as $\sum_{v\in V}(|L_{in}(v)|+|L_{out}(v)|)$) while satisfying the cover property is NP-hard \cite{2hub}. Thus, existing methods \cite{2hub,hhl,hlopt,landmark-labeling} are all heuristic to approximate the minimal 2-hop labeling index. Alternatively, we may use an all-pairs shortest path algorithm to generate index. Although it works, it requires index size of $O(|V|^2)$, which is not acceptable for large graphs.

\begin{example}
For the directed weighted graph in Figure \ref{kssr-motivation}, a possible 2-hop label indexes $L_{in}$ and $L_{out}$ is shown in Table \ref{label_index}. Suppose we compute the least cost from vertex $a$ to vertex $c$, i.e., $dis(a,c)$, we look up $L_{out}(a)$ and $L_{in}(c)$, and find the matching label entries $(s,10), (t,12)$ in $L_{out}(a)$ and $(s,10), (t,15)$ in $L_{in}(c)$, respectively. Since $10+10=20<12+15=27$, we return 20 as the result of $dis(a,c)$.
\end{example}


\begin{table}[t]
  \vspace*{-0.15in}
  \centering
  \small
  \setlength{\tabcolsep}{1pt}
  \caption{A label index for Fig. \ref{kssr-motivation}}
    \begin{tabular}{c|c|c} \hline
    \textbf{Vertex} & \textbf{$L_{in}(v)$} & \textbf{$L_{out}(v)$}\\ \hline
    $a$ & \makecell[tl]{$(a,0),(s,8),(t,33)$}& \makecell[tl]{$(a,0),(b,5),(e,6),(s,10)$,\\$(t,12)$}  \\
    $b$ & \makecell[tl]{$(b,0),(s,13),(t,20)$}& \makecell[tl]{$(b,0),(s,5),(t,7)$}\\
    $c$ & \makecell[tl]{$(c,0),(s,10),(t,15)$}& \makecell[tl]{$(b,5),(c,0),(d,3),(s,10)$,\\$(t,7)$}\\
    $d$ & \makecell[tl]{$(b,3),(d,0),(e,3),(s,13)$,\\$(t,13)$}& \makecell[tl]{$(d,0),(t,4)$}\\
    $e$ & \makecell[tl]{$(e,0),(s,14),(t,10)$}& \makecell[tl]{$(e,0),(t,7)$}\\
    $f$ & \makecell[tl]{$(e,10),(f,0),(s,24),(t,20)$}& \makecell[tl]{$(f,0),(t,3)$}\\
    $s$ & \makecell[tl]{$(s,0),(t,25)$}& \makecell[tl]{$(s,0),(t,17)$}\\
    $t$ & \makecell[tl]{$(t,0)$}& \makecell[tl]{$(t,0)$}\\ \hline
    \end{tabular}
  \label{label_index}
  \vspace*{-0.1in}
\end{table}

\begin{table}[t]
  \centering
  \small
  \caption{The inverted label index of category $\mathit{MA}$, $IL(\mathit{MA})$}
    \begin{tabular}{cc} \hline
    \textbf{Inverted label $IL(v)$} & \textbf{Label entries}\\ \hline
    $IL(a)$ & \makecell[tl]{$(a,0)$}\\
    $IL(c)$ & \makecell[tl]{$(c,0)$}\\
    $IL(s)$ & \makecell[tl]{$(a,8),(c,10)$}\\
    $IL(t)$ & \makecell[tl]{$(c,15),(a,33)$}\\ \hline
    \end{tabular}
  \label{inverted_index}
  \vspace*{-0.5cm}
\end{table}

By using the label index, an easy way to find the $x$-th nearest neighbor of $v_i$ in category $C_{i+1}$ is, for each $u\in V_{C_{i+1}}$, compute $dis(v_i,u)$ by looking up $L_{out}(v_i)$ and $L_{in}(u)$. By maintaining a min heap of size $x$, the $x$-th nearest neighbor of $v_i$ is the vertex $u$ with the $x$-th least $dis(v_i,u)$ among all vertices in $V_{C_{i+1}}$. Therefore, the time complexity is $O(\Sigma_{u\in V_{C_{i+1}}}(|L_{out}(v_i)|+|L_{in}(u)|)+|C_{i+1}|\log x)$, which is inefficient for categories with many vertices in large graphs. To improve the efficiency of \textsf{FindNN}, we construct an \emph{inverted label index} for each category, so that we can quickly identify the matching label entries between $v_i$ and all vertices in $C_{i+1}$.

The inverted label index for a category $C_i$, denoted as $IL(C_i)$, consists of label elements $IL(u')$, where $u'\in V$ is the vertex in the label entry belongs to $L_{in}(u)$ for each $u\in V_{C_i}$. That is, $IL(u')$ consists of a list of label entries $(u,d_{u',u})$, such that $u\in V_{C_i}$ and $(u',d_{u',u})\in L_{in}(u)$, and all label entries in $IL(u')$ are sorted by their costs, i.e., $d_{u',*}$, in an ascending order. With the inverted label index $IL(C_i)$, for each label entry $(u',d_{v,u'})\in L_{out}(v)$, the vertices with matching label entry in $C_i$ can be found in $IL(u') \in IL(C_i)$. Since the label entries in the inverted label index are sorted, to find the $x$-th nearest neighbor of $v$ in category $C_i$, only one label entry in each $IL(u')$ needs to be checked. 

\begin{example} \label{firstnn}
Table \ref{inverted_index} shows the inverted label index of category $\mathit{MA}$ with respect to the label index in Table \ref{label_index}. Let's find the nearest neighbor of $s$ in category $\mathit{MA}$. Since $L_{out}(s)=\{(s,0),(t,17)\}$, we look up $IL(s)$ and $IL(t)$ in $IL(\mathit{MA})$. Because label entries are sorted, only $(a,8)$ in $IL(s)$ and $(c,15)$ in $IL(t)$ need to be considered, then the nearest neighbor of $s$ in $\mathit{MA}$ is $a$ with the cost of $0+8=8$.
\end{example}


Based on the inverted label index, the detail process of finding the $x$-th nearest neighbor of $v_i$ in category $C_{i+1}$ is described by Algorithm \ref{findnn}. To avoid overlapping search, we maintain an array list $\mathit{NL}$ for $v_i$ to keep its nearest neighbors that have been found. Moreover, to avoid searching from scratch every time, we keep the candidate label entries $(u',d_{v',u'})$ in matching inverted label $IL(v')$ such that $(v',d_{v_i,v'})\in L_{out}(v_i)$ that have been found so far into a priority queue $\mathit{NQ}$ and all entries are sorted by $d_{v_i,v'}+d_{v',u'}$ in an ascending order. In addition, to keep the entry position that we have scanned for each $IL(v')$, we introduce a hash table structure $\mathit{KV}$, where $key$ is vertex $v'$ and $value$ is the entry position of $IL(v')$. $\mathit{NL}$, $\mathit{NQ}$ and $\mathit{KV}$ are all global variables and initialized to be empty. By using above data structures, to find the $x$-th nearest neighbor of $v_i$ in $C_{i+1}$, we can start from last nearest neighbor searching instead of finding the top-$x$ nearest neighbors from scratch, so that no overlapping search is needed. Specifically, if the $x$-th nearest neighbor is in $\mathit{NL}$, it can be retrieved and returned (lines 4--5). Otherwise, for the first time to find the 1st nearest neighbor of $v_i$, we retrieve all the matching inverted labels $IL(v')\in IL(C_{i+1})$, then insert the first label entry of each $IL(v')$ into $\mathit{NQ}$ and initialize $\mathit{KV}$ (lines 6--10). Subsequently, we get the minimal label entry $(u,d_{v',u})$ in $\mathit{NQ}$ which is the next nearest neighbor (line 11). In addition, we add the next label in $IL(v')$ into $\mathit{NQ}$ and update its entry position for latter nearest neighbor search (lines 12--16). Since nearest neighbors are incrementally needed, the next nearest neighbor will be the $x$-th nearest neighbor, so we add it to $\mathit{NL}$ and return (lines 17--18).

\begin{algorithm}[t!]
\caption{$FindNN(v_i,C_{i+1},x)$ 
}
\label{findnn}

\KwIn {
 Vertex $v_i$, category $C_{i+1}$, integer $x$.}

\KwOut {
The $x$-th nearest neighbor of $v_i$ in $C_{i+1}$.}

\State {$\mathit{NL} \leftarrow$ list of $v_i$'s neighbors in $C_{i+1}$ that have been found;}
\State {$\mathit{NQ} \leftarrow$ priority queue of $v_i$ for the label entries in $IL(v)\in IL(C_{i+1})$;}
\State {$\mathit{KV} \leftarrow$  $v_i$'s hash table structure for $IL(v)\in IL(C_{i+1})$;}
\If{$|\mathit{NL}|\ge x$} {
    \Return {$\mathit{NL}[x]$;}
}
\If{$|\mathit{NL}|=0$}{
    \For {each label entry $(v',d_{v_i,v'})\in L_{out}(v_i)$}{
        \State {$(u',d_{v',u'})\leftarrow IL(v')[1]$;}
        \State {$\mathit{NQ}.\textsf{insert}((u',d_{v',u'}))$;}
        \State {$\mathit{KV}.\textsf{add}(v',1)$;}
    }
}
\State {$(u,d_{v',u})\leftarrow \mathit{NQ}.\textsf{extractMin}()$;}
\Do {$u'\notin \mathit{NL}$}{
\State {$\mathit{KV}.\textsf{add}(v', \mathit{KV}.\textsf{get}(v')+1)$;}
\State {$(u',d_{v',u'})\leftarrow IL(v')[\mathit{KV}.\textsf{get}(v')]$;}

}
\State {$\mathit{NQ}.\textsf{insert}((u',d_{v',u'}))$;}
\State {$\mathit{NL}.\textsf{add}(u)$;}
\Return {$u$;}
\end{algorithm}

\begin{example}
Consider the inverted label index in Table \ref{inverted_index}, we find the 2nd nearest neighbor of $s$ in category $\mathit{MA}$. Let's follow Example \ref{firstnn}, since the 1st nearest neighbor of $s$ is $a$ from $IL(s)$, after finding $a$, for $s$, $\mathit{NL}=\{a\}$, $\mathit{NQ}=\{(c,10),(c,15)\}$ and $\mathit{KV}=\{\langle s,2\rangle,\langle t,1\rangle\}$. Hence, we get the minimal label $(c,10)$ in $\mathit{NQ}$. Because all the labels in $IL(s)$ are scanned, we set the entry position of $IL(s)$ to `-'. At this point, $\mathit{KV}=\{\langle s,-\rangle,\langle t,1\rangle\}$, $\mathit{NL}=\{a,c\}$ and $\mathit{NQ}=\{(c,15)\}$. We return $c$ as the 2nd nearest neighbor of $s$ with the cost of $0+10=10$.
\end{example}

\begin{sloppypar}
After the inverted label index is constructed offline, finding the 1st nearest neighbor of $v$ takes $O(|L_{out}(v)|\log |L_{out}(v)|)$ time, because it scans all label entries in $L_{out}(v)$ and adds the first label entry of the matching inverted label to the priority queue, and it only takes $O(\log |L_{out}(v)|)$ time to find the next nearest neighbors, which is very efficient. Let's reconsider Lemma \ref{complexity}, suppose the average index size of $L_{out}(v)$ for all $v\in V$ is $|L_{out}|$, the expected complexity of Algorithm \ref{pruningkssr} will be $O(N|L_{out}|\log |L_{out}|+(M-N)\log |L_{out}|+M\log N)$.
\end{sloppypar}

Given a witness that we have found, to get the corresponding actual route, we need to restore the route between consecutive vertices in the witness. By adding a parent vertex in each label entry of the hop labeling, it is easy to construct the actual route between two vertices \cite{landmark-labeling}. Hence, the actual route can be restored by concatenating all sub-routes between consecutive vertices in the witness.

\subsection{Integrating A$^*$ Heuristic Estimation} \label{targetmethod}
Inspired by A$^*$ algorithm \cite{astar}, the efficiency of KOSR can be further improved by using a destination-based strategy. To quickly find the feasible route, the  partially explored candidate routes with a smaller cost but far away from the destination should be given lower priority to be examined, so that the number of candidate routes can be reduced. To this end, for each  partially explored candidate route $p$, we heuristically estimate the cost of the optimal feasible route extended from $p$, so that we can examine routes according to their estimated costs instead of their real costs in an A$^*$ manner.

\begin{sloppypar}
Given a KOSR query $\langle s,t,C,k\rangle$, for a partially explored candidate route (witness) $p=\langle v_0=s,v_1,\cdots,v_i\rangle$, the optimal feasible route extended from $p$ can be represented as $p'=\langle s,v_1,\cdots,v_i,v_{i+1}, \cdots,v_{|C|},t\rangle$, so that $w(p')=w(p)+w(\langle v_i,v_{i+1},\cdots,v_{|C|},t\rangle)$. That is, we need to estimate the cost of $\langle v_i,v_{i+1},\cdots,v_{|C|},t\rangle$ which is the optimal sequenced route starts from $v_i$ and passes through $p$'s remaining categories and reaches the destination $t$. We say that a heuristic estimation $h$ for a route $P$ is \emph{admissible} if $h(P)\le w(P)$. Recall that $dis(u,v)$ returns the least cost from vertex $u$ to vertex $v$ along all possible routes from $u$ to $v$, and it can be easily computed by 2-hop labeling. Thus, we have $dis(v_i,t)\le w(\langle v_i,v_{i+1},\cdots,v_{|C|},t\rangle)$, which means $dis(v_i,t)$ is an admissible estimation of the cost of route $\langle v_i,v_{i+1},\cdots,v_{|C|},t\rangle$. Therefore, the estimated cost of $p'$ is $w(p)+dis(v_i,t)$. By applying this target-directed estimation, we propose another improved method \emph{StarKOSR}. Instead of ordering the routes, i.e., $p=\langle v_0,v_1,\cdots,v_i\rangle$, in the priority queues ($Q$ and priority queues in $\textbf{HT}_{\succ_C}$) by their real costs, i.e., $w(p)$, in StarKOSR, we order routes by their estimated costs, i.e., $w(p)+dis(v_i,t)$, so that the optimal feasible routes can be progressively found.
\end{sloppypar}

The detail process of StarKOSR is almost the same as Algorithm \ref{pruningkssr} except for \textsf{FindNN}. Since we examine routes by their estimated costs, instead of finding the $x$-th nearest neighbor of vertex $v_i$, we find $v_i$'s neighbor $v_{i+1}$ in category $C_{i+1}$ such that $dis(v_i,v_{i+1})+dis(v_{i+1},t)$ is the $x$-th least among all vertices in $C_{i+1}$, we call $v_{i+1}$ the $x$-th \emph{nearest estimated neighbor} of $v_i$. To this end, we devise the algorithm \textsf{FindNEN} (Algorithm \ref{estimate}).

\begin{algorithm}[t]
\caption{$FindNEN(v_i,C_{i+1},x)$ 
}
\label{estimate}

\KwIn {
 Vertex $v_i$, category $C_{i+1}$, integer $x$.}

\KwOut {
Vertex $v_{i+1}$ in $C_{i+1}$ such that $dis(v_i,v_{i+1})+dis(v_{i+1},t)$ is the $x$-th least.}

\State {$\mathit{ENL} \leftarrow$ list of $v_i$'s estimated neighbors in $C_{i+1}$ that have been found;}
\State {$\mathit{ENQ} \leftarrow$ priority queue of $v_i$ for the candidate neighbors in $C_{i+1}$;}
\State {$\mathit{ln} \leftarrow$  $v_i$'s last nearest neighbors that have been computed in $C_{i+1}$;}
\If{$|\mathit{ENL}|\ge x$} {
    \Return {$\mathit{ENL}[x]$;}
}
\While {$(|\mathit{ENL}|=0\bigwedge |\mathit{ENQ}|=0) \bigvee (ln\ne NULL \bigwedge dis(v_i,\mathit{ln})<dis(v_i,v)+dis(v,t))$, where $v$ is the vertex in $\mathit{ENQ}$ with minimal $dis(v_i,v)+dis(v,t)$}{
    \If {$\mathit{ln} \neq NULL$} {
        \State {$\mathit{ENQ}.\textsf{insert}(\mathit{ln})$;}
    }
    \State {$\mathit{ln}\leftarrow \textsf{FindNN}(v_i,C_{i+1},|\mathit{ENL}|+|\mathit{ENQ}|+1)$;}
}
\State {$v_{i+1}\leftarrow \mathit{ENQ}.\textsf{extractMin}()$;}
\State {$\mathit{ENL}.\textsf{add}(v_{i+1})$;}

\Return {$v_{i+1}$;}
\end{algorithm}

Given vertex $v_i$, category $C_{i+1}$ and integer $x$, Algorithm \ref{estimate} finds $v_i$'s $x$-th nearest estimated neighbor, $v_{i+1}$, in $C_{i+1}$. To avoid computing $v_i$'s $x$-th nearest estimated neighbor multiple times, we maintain an array list $\mathit{ENL}$ of $v_i$ to keep the nearest estimated neighbors that have been computed. Moreover, to continuously compute the next nearest estimated neighbors, we maintain a priority queue $\mathit{ENQ}$ of $v_i$ for candidate neighbors that have been considered so far and sort the neighbors ($v$) by their estimated costs ($dis(v_i,v)+dis(v,t)$) in an ascending order. Meanwhile, we store the last nearest neighbor of $v_i$ that have been computed into variable $\mathit{ln}$, so that we can start from last nearest estimated neighbor searching instead of computing from scratch every time. We note that $\mathit{ENL}$, $\mathit{ENQ}$ and $\mathit{ln}$ are global variables and initialized to be empty or NULL. Then if the $x$-th nearest estimated neighbor of $v_i$ has been computed, we can retrieve it from $\mathit{ENL}$ instead of recomputing it (lines 4--5). Otherwise, we find the $x$-th nearest estimated neighbor for the first time. Instead of checking all vertices in $C_{i+1}$ to find the $x$-th nearest estimated neighbor, we incrementally find the next nearest neighbor $\mathit{ln}$ of $v_i$ in $C_{i+1}$ by calling \textsf{FindNN} (line 9), if $dis(v_i,\mathit{ln})$ is greater than the minimal cost $dis(v_i,v)+dis(v,t)$ in $\mathit{ENQ}$, then $v$ has the minimal estimated cost among all remaining vertices in $C_{i+1}$, because other vertices, say $v'$, that have not been checked hold $dis(v_i,v')\ge dis(v_i,\mathit{ln})\ge dis(v_i,v)+dis(v,t)$, which means that their estimated cost cannot be less than that of $v$. Since \textsf{FindNEN} is incrementally called, the next nearest estimated neighbor is the $x$-th nearest estimated neighbor. Finally, we add $v$ to $\mathit{ENL}$ and return it as the result (lines 10--12). 

\begin{lemma}
Algorithm StarKOSR returns the correct result for a KOSR query $(s, t, C=\langle C_1,\cdots,C_j\rangle, k)$.
\end{lemma}

\begin{proof}
  Suppose the examined route from the priority queue at each iteration is $P=\langle s,v_1,\cdots,v_i\rangle$ $(1\le i\le |C|$ and $v_j\in V_{C_j}$ for $1\le j\le i)$, we first prove the total estimated cost of $P$, i.e., $w(P)+dis(v_i,t)$, is minimal in all possible partially explored routes. For all partially explored routes in the priority queue, it is trivial that they have greater total estimated costs than $P$. For the possible partially explored routes that have not been generated, they must have greater total estimated costs than the routes in the priority queue, since we always generate nearest estimated neighbor for each examined routes. Next, we prove that $P$ is the optimal route from $s$ to $v_i$ and passes through category sequence $\langle C_1,\cdots,C_i\rangle$. Suppose there is a route $P'=\langle s,v'_1,\cdots,v'_q\rangle$ $(q<i)$ such that $w(P')+w(\langle v'_q,\cdots,v_i\rangle)<w(P)$, then we have $w(P')+dis(v'_q,t)\le w(P')+w(\langle v'_q,\cdots,v_i\rangle)+dis(v_i,t)< w(P)+dis(v_i,t)$, which means that the total estimated cost of $P'$ is smaller than $P$, which is a contradiction. Thus, $P$ is the optimal route from $s$ to $v_i$ and passes through category sequence $\langle C_1,\cdots,C_i\rangle$. Hence, when $v_i$ equals to destination $t$, we have found an optimal route from $s$ to $t$ and passes through category sequence $C$. In summary, StarKOSR returns the correct result for a KOSR query.
\end{proof}

\begin{example} \label{star}
Reconsider Figure \ref{kssr-motivation} and suppose the KOSR query is $(s,t,\langle \mathit{MA},\mathit{RE},\mathit{CI}\rangle,2)$. Table \ref{star_example} lists the routes in the priority queue $Q$ at each step by applying StarKOSR algorithm. The first route is $\langle s\rangle$, and we extend $\langle s\rangle$ by finding the 1st nearest estimated neighbor of $s$ in category $\mathit{MA}$. Initially, the $\mathit{ln}$ of $s$ is $NULL$, $\mathit{ENQ}$ and $\mathit{ENL}$ are empty, we find the 1st nearest neighbor of $s$ which is $a$ with $dis(s,a)+dis(a,t)=8+12=20$, then we add $a$ to $\mathit{ENQ}$ and continue to find the 2nd nearest neighbor $c$ with $dis(s,c)+dis(c,t)=10+7=17$. Because $dis(s,c)=10<dis(s,a)+dis(a,t)=8+12=20$, $c$ is added to $\mathit{ENQ}$. Since $s$ has no next nearest neighbor in $\mathit{MA}$, $ln=NULL$ and $c$ is the 1st nearest estimated neighbor of $a$ (step 2), thus, we have $\mathit{ln}=NULL$, $\mathit{ENQ}=\{a\}$ and $\mathit{ENL}=\{c\}$ after $c$ is returned. Subsequently, we extend $\langle s,c\rangle$ via $b$ and generate candidate route $\langle s,a\rangle$ by finding the 2nd nearest estimated neighbor of $s$ in category $\mathit{MA}$ (step 3). At step 6, the first optimal sequenced route $\langle s,a,b,d,t\rangle$ is found and no dominated routes exist. Finally, the second optimal feasible route $\langle s,a,e,d,t\rangle$ is found at step 9.
\end{example}

\begin{table}[t]
  \vspace*{-0.1in}
  \centering
  \small
  \setlength{\tabcolsep}{1pt}
  \caption{Running example of StarKOSR for Fig. \ref{kssr-motivation}}
    \begin{tabular}{cc} \hline
    \textbf{Step} & \textbf{Routes (route(estimated cost), x) in priority queue $Q$}\\ \hline
    1 & \makecell[tl]{$(\langle s\rangle(0),1)$}  \\
    2 & \makecell[tl]{$(\langle s,c\rangle(17),1)$} \\
    3 & \makecell[tl]{$(\langle s,a\rangle(20),2)$,$(\langle s,c,b\rangle(22),1)$}\\
    4 & \makecell[tl]{$(\langle s,a,b\rangle(20),1)$,$(\langle s,c,b\rangle(22),1)$}\\
    5 & \makecell[tl]{$(\langle s,a,b,d\rangle(20),1)$,$(\langle s,a,e\rangle(21),2)$,$(\langle s,c,b\rangle(22),1)$}\\
    6 & \makecell[tl]{\underline{$(\langle s,a,b,d,t\rangle(20),1)$},$(\langle s,a,e\rangle(21),2)$,$(\langle s,c,b\rangle(22),1)$,\\$(\langle s,a,b,f\rangle(43),2)$}\\
    7 & \makecell[tl]{$(\langle s,a,e\rangle(21),2)$,$(\langle s,c,b\rangle(22),1)$,$(\langle s,a,b,f\rangle(43),2)$}\\
    8 & \makecell[tl]{$(\langle s,a,e,d\rangle(21),1)$,$(\langle s,c,b\rangle(22),1)$,$(\langle s,a,b,f\rangle(43),2)$}\\
    9 & \makecell[tl]{\underline{$(\langle s,a,e,d,t\rangle(21),2)$},$(\langle s,c,b\rangle(22),1)$,$(\langle s,a,e,f\rangle(27),2)$,\\$(\langle s,a,b,f\rangle(43),2)$}\\ \hline
    \end{tabular}
  \label{star_example}
  \vspace*{-0.5cm}
\end{table}

As we can see from Example \ref{star}, 4 steps are reduced compared to Example \ref{prune_example}. That is, StarKOSR further reduces the searching space and improves the efficiency of KOSR.
In StarKOSR, though \textsf{FindNN} may be called multiple times as we attempt to find the $x$-th nearest estimated neighbor by applying \textsf{FindNEN}, however, to find the next optimal feasible route, the total times of calling \textsf{FindNN} by StarKOSR is significantly less than that by PruningKOSR. We address this as follows: suppose we examine route $p=\langle v_0,\cdots,v_i\rangle$, and find the $x$-th nearest estimated neighbor $v_{i+1}$ of $v_i$ by calling \textsf{FindNN} $j$ times, that is $j$ nearest neighbors of $v_i$ have been found and $dis(v_i,NN_y)<dis(v_i,v_{i+1})+dis(v_{i+1},t)$ for each nearest neighbor $NN_y, 1\le y<j$. If $w(p)+dis(v_i,v_{i+1})+dis(v_{i+1},t)<w(P)$, where $w(P)$ is the cost of the next optimal feasible route $P$, then $w(p)+dis(v_i,NN_y)<w(P)$, that is, to find $P$, $j-1$ candidate routes $\langle v_0,\cdots,v_i,NN_y\rangle$ should be examined in PrunningKOSR by calling \textsf{FindNN} $j$ times. In this case, both methods call \textsf{FindNN} the same times. On the other hand, if $w(p)+dis(v_i,v_{i+1})+dis(v_{i+1},t)\ge w(P)$, then $\langle v_0,\cdots,v_i,v_{i+1}\rangle$ won't be examined and subsequently, all possible candidate routes derived from $\langle v_0,\cdots,v_i,v_{i+1}\rangle$ can never be considered before $P$ is found, which in turn reduces the searching space of StarKOSR.  Thus, in summary, the times of calling \textsf{FindNN} by StarKOSR is significantly less than that by PruningKOSR.

\begin{figure}[t!]
\vspace*{-0.4cm}
\centering
\includegraphics[width=3.5in]{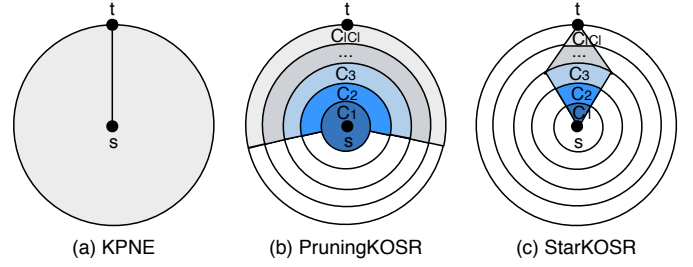}
\caption{The searching space of different methods}
\label{spaceF}
\vspace*{-0.4cm}
\end{figure}

\stitle{Remarks.} Figure \ref{spaceF} illustrates the searching space of different methods for the first optimal sequenced route. Since KPNE examines all possible candidate routes that with smaller costs than the optimal sequenced route, its searching space (Figure \ref{spaceF}(a)) is a whole circle whose radius is the cost of the optimal sequenced route from source $s$ to destination $t$, and each route in the circle will be examined.

In PruningKOSR, for each category $C_i$, at most $|C_i|$ routes are extended due to dominance relations, which results in at most $|C_i|\cdot|C_{i+1}|$ candidate routes can be examined at category $|C_{i+1}|$, which is the area of each dark ring in Figure \ref{spaceF}(b). As a result, the searching space (area) of PruningKOSR is reduced compared to KPNE, and the pruned space consists of the routes that are dominated and the candidate routes derived from them.

For StarKOSR, since we consider the whole cost of the route from source to destination by using target-directed strategy, the  partially explored candidate routes that are far away from the destination are further pruned, as a result, the area of each ring in Figure \ref{spaceF}(c) gets smaller compared to Figure \ref{spaceF}(b). Since the estimated whole costs of the  partially explored routes are not greater than the real costs of their corresponding optimal sequenced routes, and as we extend routes along the category sequence, the estimated whole costs become larger and closer to the real optimal cost. As a result, at the beginning, loose estimated cost (may not contain the required categories) enables more candidate routes to be examined and the searching space (area) increases. Subsequently, when the estimated costs get tighter and are closer and closer to the real optimal cost and finally equal to the real optimal cost, more and more routes whose estimated costs are greater than the optimal cost are filtered and the searching space (area) shrinks until the optimal sequenced route is found.


\subsection{Extensions} \label{extensions}
In this section, we extend our work in several aspects.

\stitle{Variants of KOSR:} The KOSR query can have different variants, which can be solved by extending our proposal. For KOSR on \emph{unweighted graphs}, we simply set the weights of all edges to 1. For \emph{undirected graphs}, for each vertex $v$, $L_{in}(v)$ and $L_{out}(v)$ are identical in the 2-hop labeling, and thus only one label is needed. In the case that source is not required, we can initially add all vertices in the first category instead of the source to the priority queue (line 3 in Algorithm 2). In the case that destination is not required, since the estimated cost to the destination cannot be applied, the StarKOSR method will not work, but PruningKOSR still works. In the case that people want to find the KOSR according to their personal preferences, for example, you may want the restaurant ($\mathit{RE}$) in your sequenced route to be an Italian restaurant, such constraint can be easily met by finding the $x$-th nearest Italian restaurant in category $\mathit{RE}$. Specifically, these constraints can be added to line 15 of Algorithm 3.

\stitle{Handling dynamic updates:} We distinguish two different kinds of updates: the graph structure updates and the category updates.

For the graph structure updates, we adopt existing methods \cite{dynamic1,dynamic2,dynamic3} to update label index. After label entries are inserted into or removed from the label index (update the cost of existing label entries can be regarded as a remove-insert operation), the corresponding inverted label indexes should also be updated accordingly.

For the category updates, which means the category set $F(v)$ of vertex $v$ has changed. If a new category $C_i$ is inserted into $F(v)$, add $v$ into $V_{C_i}$, and for each entry $(u,d_{u,v})\in L_{in}(v)$, we find the inverted label $IL(u)\in IL(C_i)$, and insert $(v,d_{u,v})$ into $IL(u)$ by using a binary search for $d_{u,v}$. On the other hand, if a category $C_i$ is removed from $F(v)$, we delete $v$ from $V_{C_i}$, and for each entry $(u,d_{u,v})\in L_{in}(v)$, find inverted label $IL(u)\in IL(C_i)$, then retrieve and remove $(v,d_{u,v})$ from $IL(u)$. Since the inverted label index is in order, above operations take $O(|L_{in}(v)|\log |C_i|)$ time, which is efficient.

\stitle{Disk-based query answering:} In the case that the label index cannot fit into memory, we store the indexes (including label indexes and inverted label indexes) into disk according to categories. Specifically, indexes in each category $C_i$ are stored as follows: $IL(C_i)$, $L_{out}(v)$ and $L_{in}(v)$ ($v\in V_{C_i}$). For each KOSR query $(s,t,C,k)$, we locate the beginning of the index of each category $C_i$ by a disk-based $B^+$ tree, and load the inverted label index $IL(C_i)$ as well as $L_{out}(v)$ for each $v\in V_{C_i}$. For source $s$ (or destination $t$), we first locate its category that it belongs to, then locate and load its label index $L_{out}(s)$ (or $L_{in}(t)$). Overall, $|C|+4$ disk seek operations are needed.

\section{Experimental evaluation} \label{sec:exp}

\subsection{Experimental Setup}

\noindent
\textbf{Datasets: } We use five real-world graphs with varying sizes. In particular, $\mathit{CAL}$, $\mathit{NYC}$, $\mathit{COL}$, and $\mathit{FLA}$ are graphs representing the road networks of California, New York City, Colorado, and Florida, respectively. $\mathit{G+}$ is the social network from $\mathit{Google+}$.
Table \ref{graphs} gives the sizes in terms of the cardinalities of both vertex and edge sets.

\begin{table}[h]
  \centering
  \small
  \vspace*{-0.1in}
  \caption{Real-World Graphs}
  \begin{tabular} {ccc} \hline
  \textbf{Dataset}  & \textbf{$|V|$} & \textbf{$|E|$}\\ \hline %
  $\mathit{CAL}$\footnotemark[1]  & 68,345& 68,990 \\ %
  $\mathit{NYC}$\footnotemark[2] & 980,632& 1,280,981 \\ %
  $\mathit{COL}$\footnotemark[3] & 435,666 & 1,057,066 \\ %
  $\mathit{FLA}$\footnotemark[3] & 1,070,376 & 2,687,902 \\ %
  $\mathit{G+}$\footnotemark[4] & 107,614 & 13,673,453 \\ \hline%
  \end{tabular}
  \label{graphs}
  \vspace*{-0.1in}
\end{table}

\footnotetext[1]{\url{http://www.cs.utah.edu/~lifeifei/SpatialDataset.htm}}
\footnotetext[2]{\url{http://www.openstreetmap.org}}
\footnotetext[3]{\url{http://www.dis.uniroma1.it/challenge9/download.shtml}}
\footnotetext[4]{\url{http://snap.stanford.edu/data/index.html}}

In particular, $\mathit{CAL}$ is a weighted, undirected graph where edge weights represent the distances of the corresponding roads. In addition, 47,298 vertices in $\mathit{CAL}$ are associated with 63 different categories. $\mathit{NYC}$ is a weighted, undirected road network downloaded from OpenStreetMap\footnotemark[2]. In addition, we also get the POI dataset of New York from OpenStreetMap\footnotemark[2]. Specifically, the POI dataset contains 30,382 points of interest in New York that belong to 135 different categories. For each POI, we find its nearest vertex in the road network and regard the category of the vertex as the category of the POI. 

Graphs $\mathit{COL}$ and $\mathit{FLA}$ are weighted directed graphs, where edge weights represent the travel time of roads. Graph $\mathit{G+}$ is an unweighted, directed graph where all edge weights are set to 1. Since no categorical information is associated with the vertices in these graphs, we generate categories for the vertices using both uniform and zipfian distributions.
In particular, we follow~\cite{gsp} to generate uniform distributions. We fix the number of vertices in each category with parameter $|C_i|$, and then uniformly assign a category to vertices. We generate uniform categories for $\mathit{COL}$, $\mathit{FLA}$, and $\mathit{G+}$, which is used as the default setting in the following experiments. Next, following~\cite{osr}, we generate 100 categories for $\mathit{FLA}$ with Zipfian distribution, and we use a parameter factor $f (\ge 1)$ to control the skewness of the distributions, the greater the $f$ is. The less skew the distributions are. For example, when $f=1.2$, the smallest category size is 23, and the largest category size is 139,717.


\noindent
\textbf{Queries: } For each KOSR query $(s,t,C,k)$, we randomly select a source-destination pair, a category sequence with size $|C|$, and an integer $k$. Then, we issue the query on all graphs. In each experiment, 50 random query instances are constructed and the average query time is reported. If a query cannot stop within 3,600 seconds, or fails due to out of memory exception, we denote its corresponding query time as \textbf{INF}.
We vary important parameters according to Table \ref{parameters}, where default parameter settings are shown in bold.

\begin{table}[h]
  \centering
  \small
  \vspace*{-0.1in}
  \caption{Parameter Settings}
  \begin{tabular} {cc} \hline
  \textbf{Parameter} & \textbf{Values} \\ \hline %
  $|C_i|$ & 5,000, \textbf{10,000}, 15,000, 20,000 \\ %
  $|C|$ & 2, 4, \textbf{6}, 8, 10 \\  %
  $k$ & 10, 20, \textbf{30}, 40, 50 \\ \hline%
  \end{tabular}
  \label{parameters}
  \vspace*{-0.12in}
\end{table}

\noindent
\textbf{Label index: } We adopt the pruned landmark labeling method \cite{landmark-labeling}, which achieves good performance and is easy to implement, to precompute the label index for each graph in Table \ref{graphs}. Based on the label index, we then construct the inverted label index for each category in the graph. Table \ref{preprocessing} shows the preprocessing results on different graphs under default parameter settings. For large graphs, e.g., $\mathit{FLA}$, the index sizes may be too large to fit into main memory. To contend with this, we store the indexes on disks. Alternatively, labeling compression method \cite{hubcom} can be applied to further reduce the index sizes.

\begin{table}[h]
  \vspace*{-0.1in}
  \centering
  \small
  \setlength{\tabcolsep}{1pt}
  \caption{Preprocessing results on different graphs}
  \begin{tabular} {ccccc} \hline
  \multicolumn{5}{c}{For label indexes}\\ \hline
  \textbf{Graph} & \textbf{Time [H:M]} & \textbf{Avg. $|L_{in}(v)|$} & \textbf{Avg. $|L_{out}(v)|$} & \textbf{Index Size}\\ \hline %
  $\mathit{CAL}$ & 0:1 & 122.90 & 122.90 & 95.24MB\\  %
  $\mathit{NYC}$ & 0:52 & 704.94 & 704.94 & 14.11GB\\  %
  $\mathit{COL}$ & 0:9 & 1,101.06 & 1,101.06 &5.05GB\\ %
  $\mathit{FLA}$ & 2:42 & 1,495.84 & 1,495.84 &18.25GB\\
  $\mathit{G+}$ & 0:2 & 335.53 & 347.96 & 230.47MB\\  \hline%
  \multicolumn{5}{c}{For inverted label indexes}\\ \hline
  \textbf{Graph} & \textbf{Time [H:M]} & \textbf{Avg. $|IL(C_i)|$} & \textbf{Avg. $|IL(v)|$} & \textbf{Index Size}\\ \hline %
  $\mathit{CAL}$ & 0:1 & 9543.48 & 13.75 & 49.05MB\\  %
  $\mathit{NYC}$ & 0:1 & 10863.24 & 12.61 & 120.50MB\\  %
  $\mathit{COL}$ & 0:3 & 98,345.80 & 121.94 &2.53GB\\ %
  $\mathit{FLA}$ & 0:14 & 181,763.43 & 82.32 &9.18GB\\  %
  $\mathit{G+}$ & 0:1 & 39,621.12 & 91.13 &113.32MB\\  \hline%
  \end{tabular}
  \label{preprocessing}
  \vspace*{-0.4cm}
\end{table}

\noindent
\textbf{Methods}: We consider the following methods for answering KOSR queries: (1) \textsf{GSP}: the state-of-the-art algorithm to find the optimal sequenced route ($k=1$). (2) \textsf{KPNE}: the KPNE algorithm (Section \ref{sub:pne}) by using Algorithm \textsf{FindNN} to find the nearest neighbors. (3) \textsf{PK}: our algorithm PruningKOSR by using dominance relationship to filter temporarily unnecessary routes (Section \ref{prunmethod}). (4) \textsf{SK}: our algorithm StarKOSR by using the target-directed strategy to find the optimal feasible routes (Section \ref{targetmethod}). (5) \textsf{SK-DB}: StartKOSR with label indexes resident on disks. (6) \textsf{KPNE-Dij, PK-Dij, SK-Dij}: the KPNE, PruningKOSR, and StarKOSR algorithms by using Dijkstra's search to find the nearest neighbors rather than using Algorithm FindNN. 

\noindent
\textbf{Evaluation Criteria}: We evaluate the performance of different methods in three different aspects: the query run-time, the number of examined routes (witnesses), and the number of (next) nearest neighbor (shorten as NN) queries executed by calling Algorithm \textsf{FindNN}, where the number of hits in the $\mathit{NL}$ list (line 5 at Algorithm \ref{findnn}) is not included.

\noindent
\textbf{Implementation details: } All algorithms are implemented in Java 1.6 and run on a Windows 10 machine with 3.2GHz CPU, and 32 GB memory.

\subsection{Experimental Results}
We first evaluate the efficiency of different  methods (except for GSP) on different graphs for answering KOSR queries under the default parameter settings, and then evaluate the effects of parameters by varying their values.
Finally, when $k=1$, we test the performance of our methods against the state-of-the-art method GSP for answering OSR queries.

\begin{figure*}[t]
\vspace*{-0.4in}
\centering
\subfigure[Query run-time]{
\label{overall}
\includegraphics[width=1.6in]{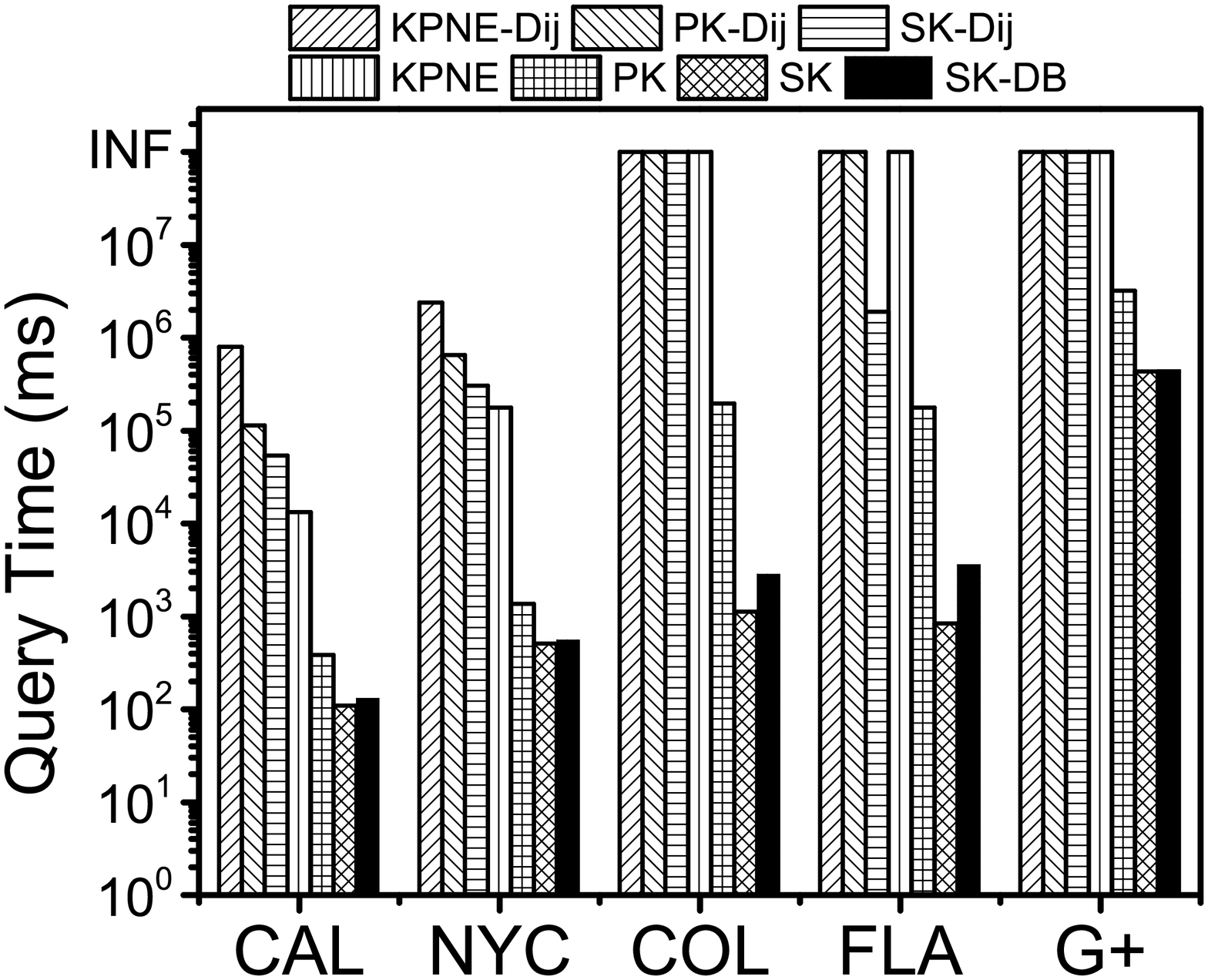}}
\subfigure[\# of examined routes]{
\label{overallpath}
\includegraphics[width=1.6in]{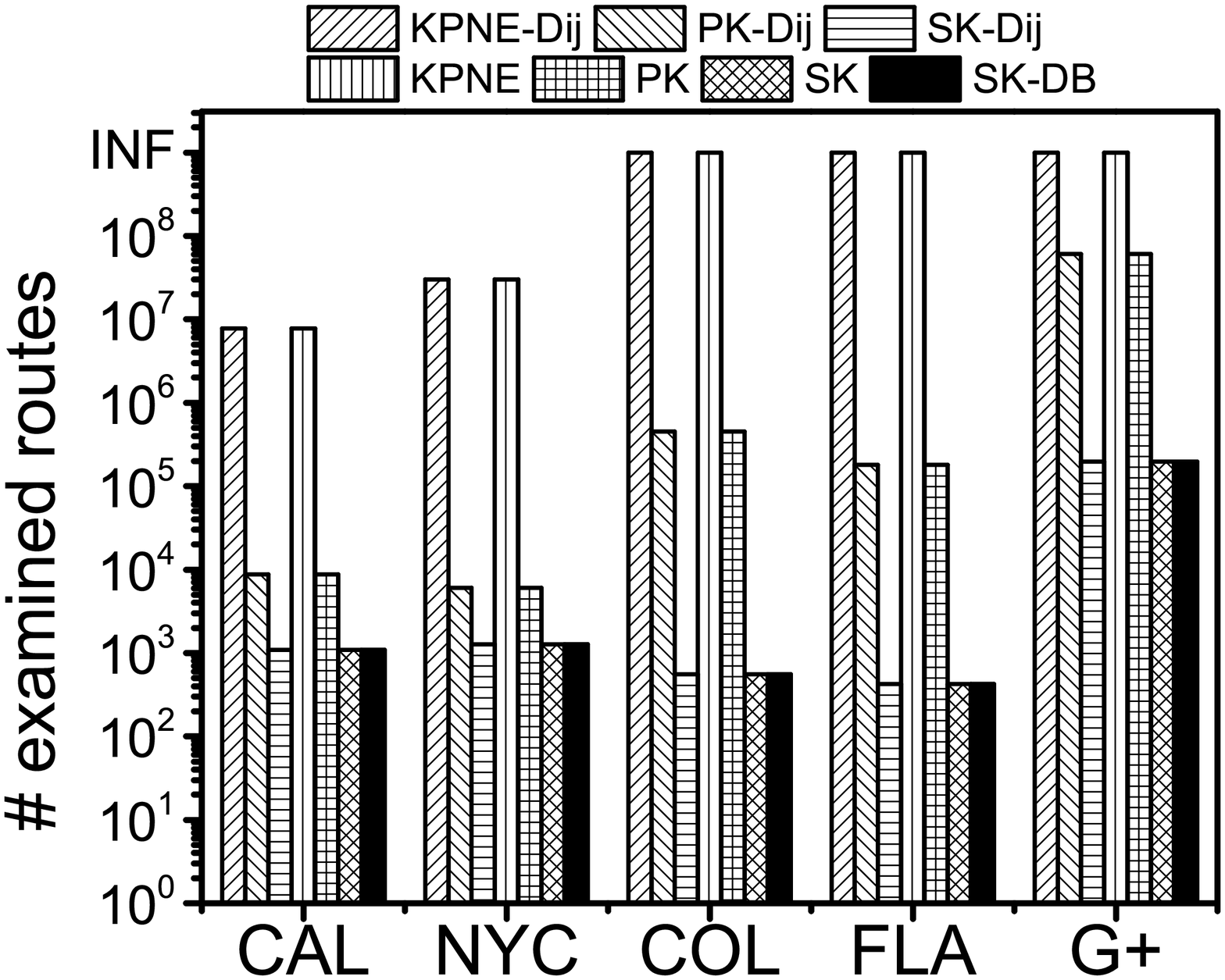}}
\subfigure[\# of NN queries]{
\label{overallnn}
\includegraphics[width=1.6in]{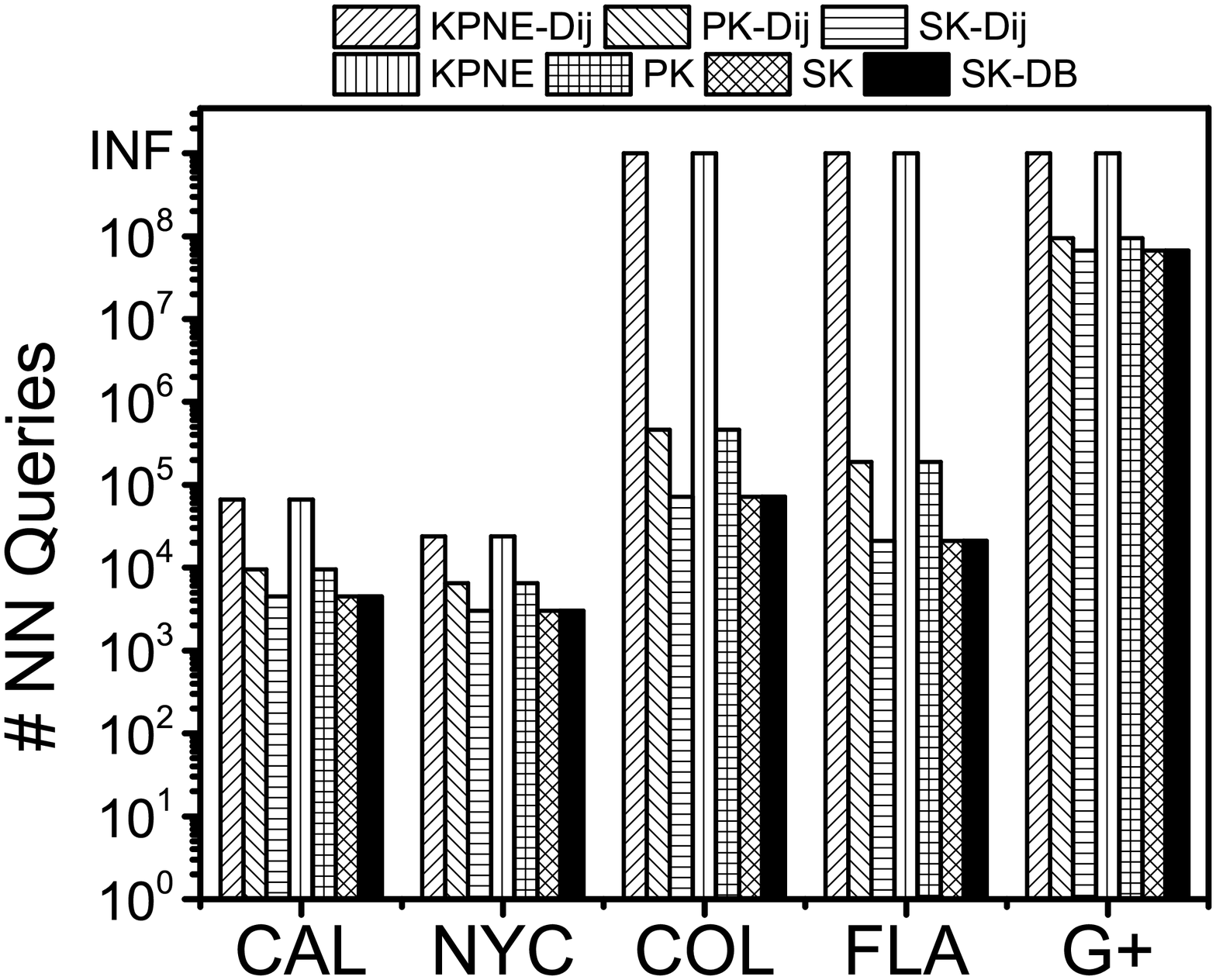}}
\subfigure[Effect of $k$, $\mathit{FLA}$]{
\label{flak}
\includegraphics[width=1.7in]{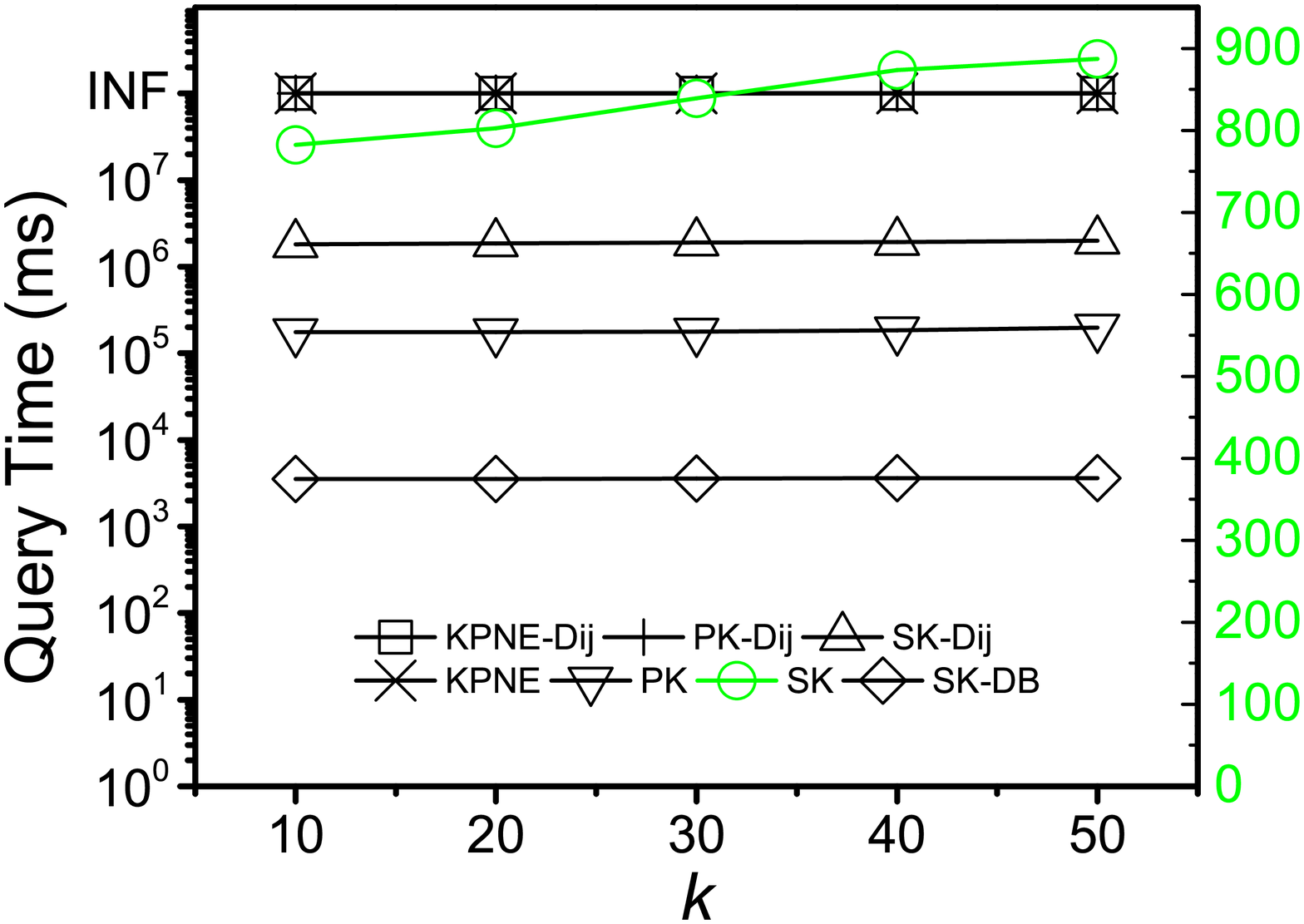}}

\subfigure[Effect of  $k$, $\mathit{CAL}$]{
\label{calk}
\includegraphics[width=1.75in]{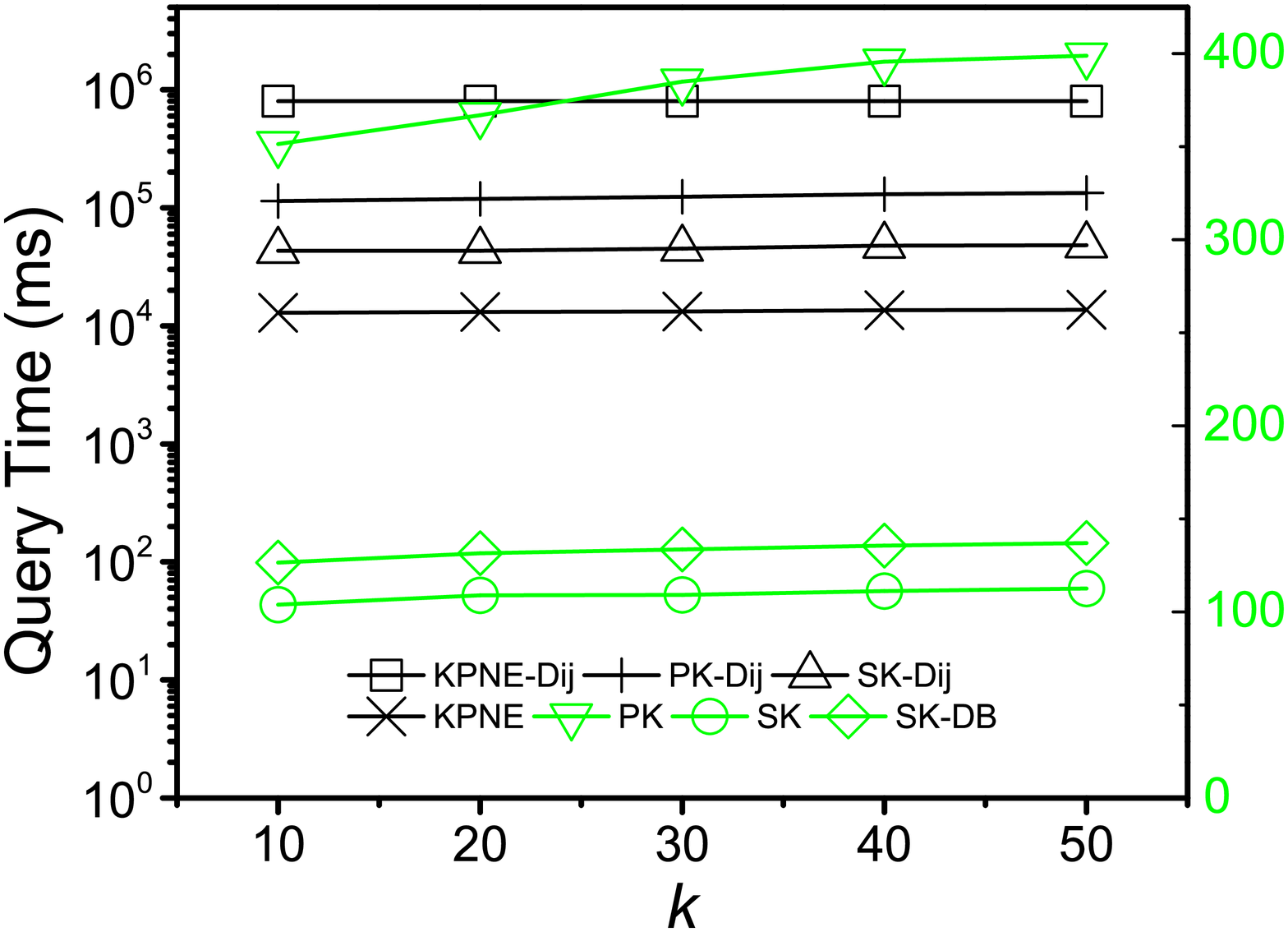}}
\subfigure[Effect of $|C|$, $\mathit{FLA}$]{
\label{flacat}
\includegraphics[width=1.6in]{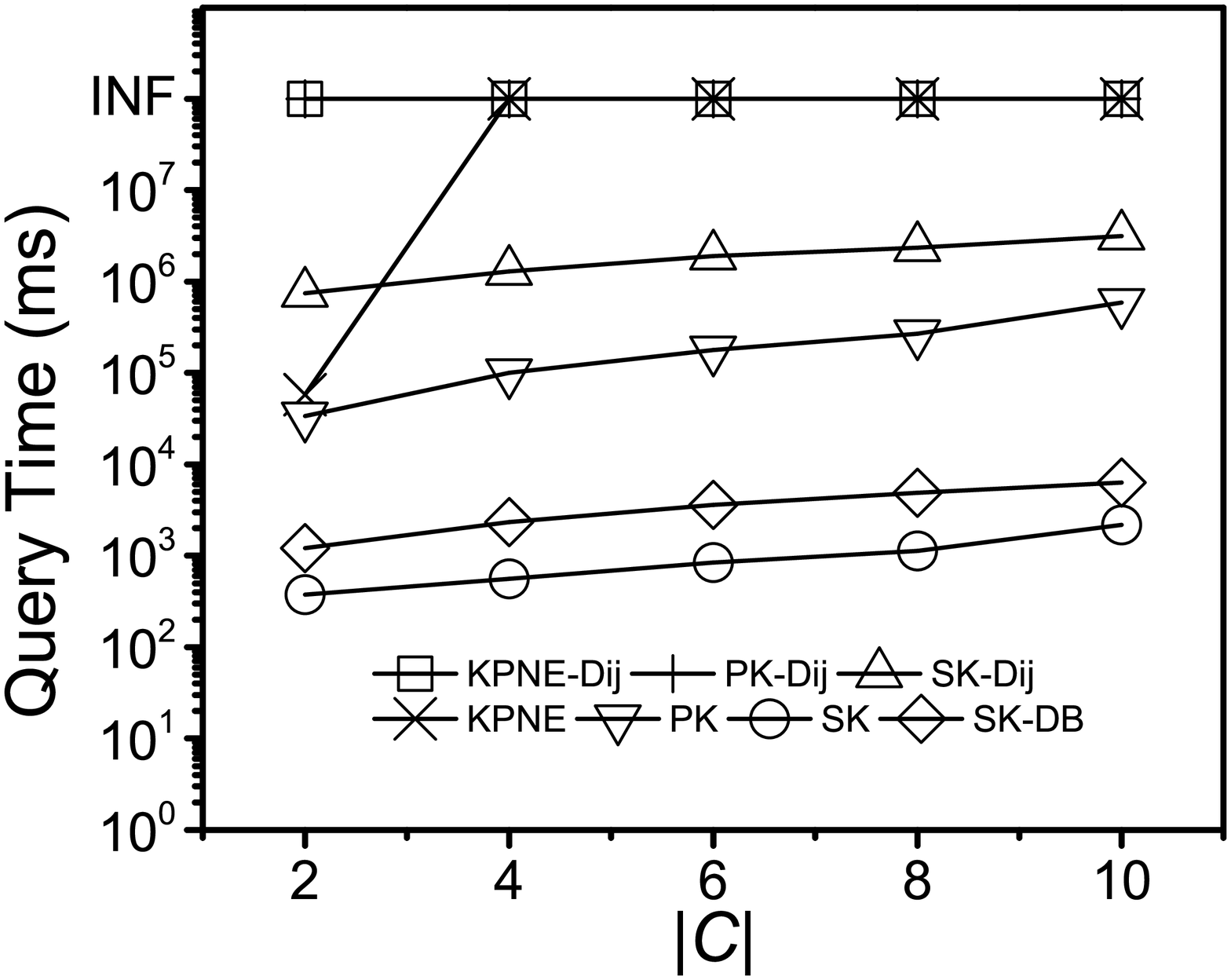}}
\subfigure[Effect of  $|C|$, $\mathit{CAL}$]{
\label{calcat}
\includegraphics[width=1.6in]{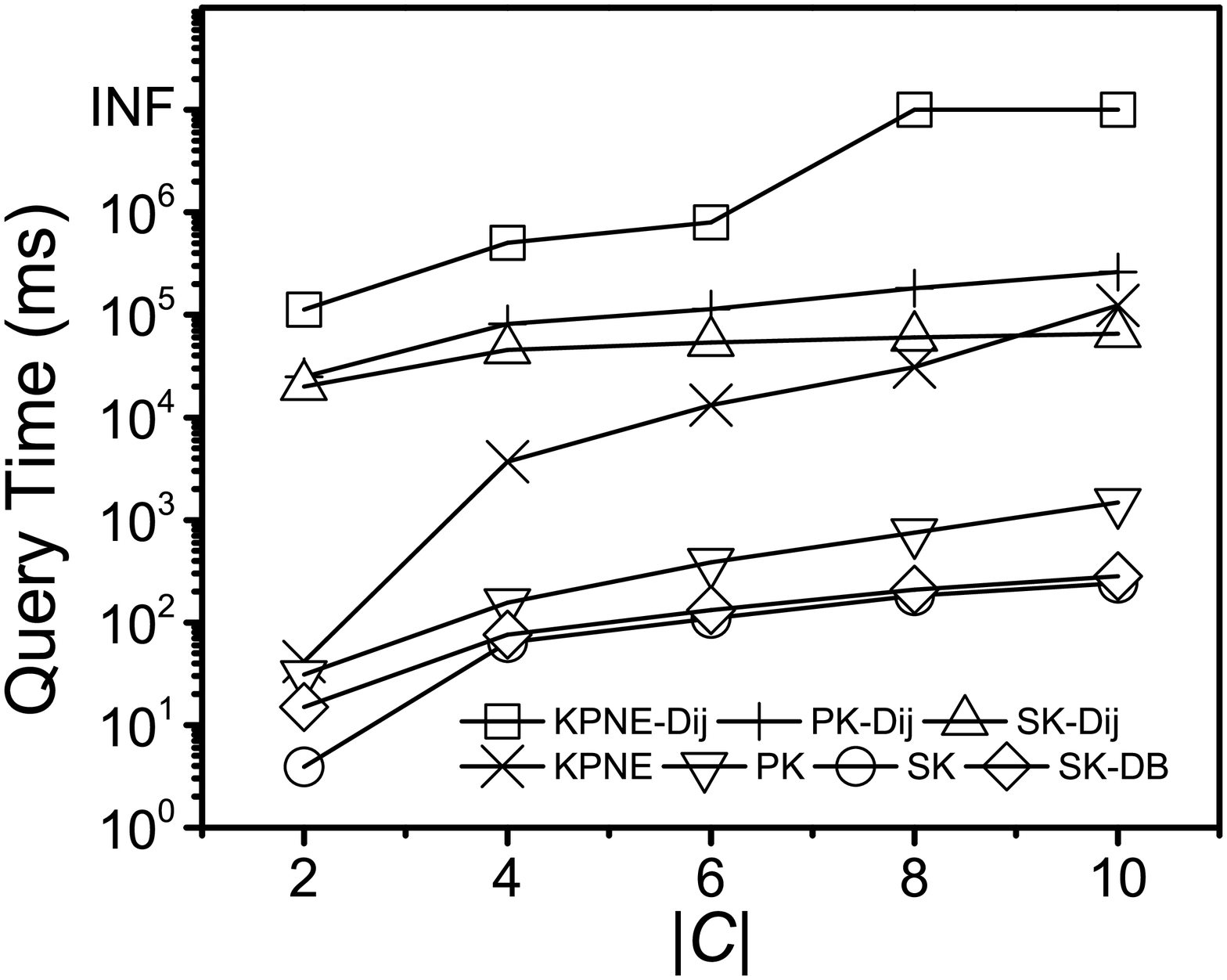}}
\subfigure[Effect of $|C_i|$, $\mathit{FLA}$]{
\label{flacap}
\includegraphics[width=1.6in]{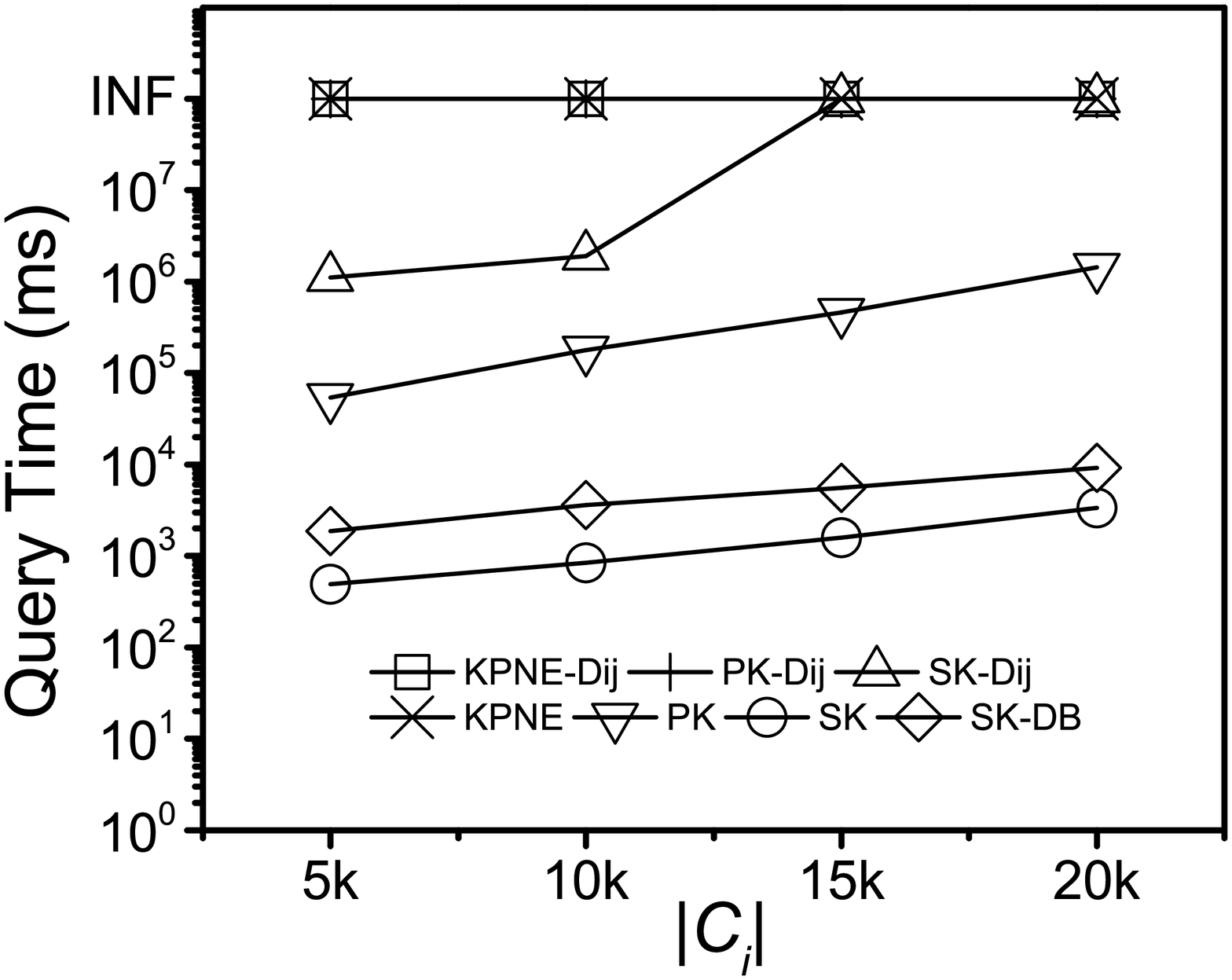}}

\caption{Performance of different methods with different parameter settings for KOSR queries}
\vspace*{-0.2in}
\end{figure*}

\noindent
\textbf{Overall performance under default parameter settings.} Figures \ref{overall}$\sim$\ref{overallnn} show the performance of different methods on different graphs. The run-times of the methods on different graphs are illustrated in Figure \ref{overall}. Since KPNE examines all possible candidate routes in the searching space, both KPNE and KPNE-Dij are not well performed and they cannot return the results on larger graphs with large category size, i.e., $\mathit{COL}$, $\mathit{FLA}$, and $\mathit{G+}$ within 3,600 seconds. Compared to KPNE, by reducing the searching space, both PK and SK are able to return the results on all graphs. Since SK further filters partially explored routes that are far away from the destination by using a target-directed cost estimation strategy, it performs nearly two orders of magnitude faster than PK on $\mathit{COL}$ and $\mathit{FLA}$, 4 (or 3) times faster than PK on $\mathit{CAL}$ (or $\mathit{NYC}$), and 7 times faster than PK on $\mathit{G+}$. In addition, by comparing the costs of the routes in the result set of different methods on $\mathit{CAL}$ and $\mathit{CAL}$, our methods, i.e., PK and SK, have the same results as KPNE, which also verifies the correctness of our methods.
On the other hand, since the time complexities of our methods are independent of the graph size, but are dependent on the size of category sequence $|C|$ and the category size $|C_i|$, both PK and SK have steady query run-times on larger graphs such as $\mathit{COL}$ and $\mathit{FLA}$. 
Moreover, PK and SK perform orders of magnitude faster than PK-Dij and SK-Dij, respectively, because Algorithm FindNN performs efficient NN queries by using inverted label indexes. Since SK-DB needs additional time to load label indexes into memory and initialize them for each query, it takes more time than SK. However, it still outperforms PK where all indexes are always resident in main memory.

For $\mathit{G+}$, since its edge weights are all 1 and its diameter is only 6, the partially explored routes and nearest neighbors tend to have similar costs, which leads to a larger searching space for both PK and SK, as a result, both PK and SK take much more time to find the top-$k$ optimal sequenced routes. Moreover, since Dijkstra's search on unweighted graph explores much more vertices and edges, all KPNE-Dij, PK-Dij and SK-Dij cannot return the results on $\mathit{G+}$.

Figure \ref{overallpath} and Figure \ref{overallnn} show the number of examined routes and NN queries, respectively, in different methods on different graphs. Clearly, the number of examined routes and NN queries in SK is much fewer than PK on all graphs, which means the searching space of SK is much smaller than that of PK. As a result, SK significantly outperforms PK. Note that the average NN queries per vertex in each examined route of SK is much greater than that of PK, for example, about 4 vs. 1 on $\mathit{CAL}$, 50 vs. 1 on $\mathit{FLA}$, and 217 vs. 1 on $\mathit{G+}$, because SK needs to compute more nearest neighbors to find the next nearest estimated neighbor. However, the total times of NN queries of SK is significantly less than that of PK, which also explains its excellent query run-time.
Note that different index loading methods (in memory vs. disk) and NN query algorithms (FindNN vs. Dijkstra's search) do not change the process of the KOSR algorithm, hence SK and SK-DB, KPNE (or PK or SK) and KPNE-Dij ( or PK-Dij or SK-Dij) have the same number of examined routes and NN queries.

\begin{table}[h]
  \centering
  \small
  \setlength{\tabcolsep}{1pt}
  \vspace*{-0.1in}
  \caption{Distributions of the query time (ms) on $\mathit{FLA}$}
  \begin{tabular} {ccc} \hline
   & \textbf{PK} & \textbf{SK}\\ \hline %
  \textbf{Overall query time} & 177,622.60 & 838.96 \\  \hline%
  \textbf{NN query time} & 177,175.84 & 732.87 \\ %
  \textbf{Priority queue maintenance time} & 303.68 & 0.11 \\
  \textbf{Estimation time} & 0 & 101.99 \\
  \textbf{Others time} & 143.08 & 3.99 \\ \hline%
  \end{tabular}
  \label{constitutions}
  \vspace*{-0.1in}
\end{table}

Table \ref{constitutions} shows the distributions of the run-times of our methods on graph $\mathit{FLA}$. Clearly, the NN queries dominate the query run-time of both methods. Since lots of candidate routes are examined in PK, the maintenance of the priority queue in PK costs more time than does SK. On the other hand, SK needs to compute the least cost to the destination to estimate the total cost for a partially explored route, which takes some time. While PK does not spend any time since it does not estimate the total cost. However, the time on cost estimation is only a small portion of the overall query time.

\begin{figure}
\begin{minipage}[t]{0.45\linewidth}
\centering
\includegraphics[width=1.6in]{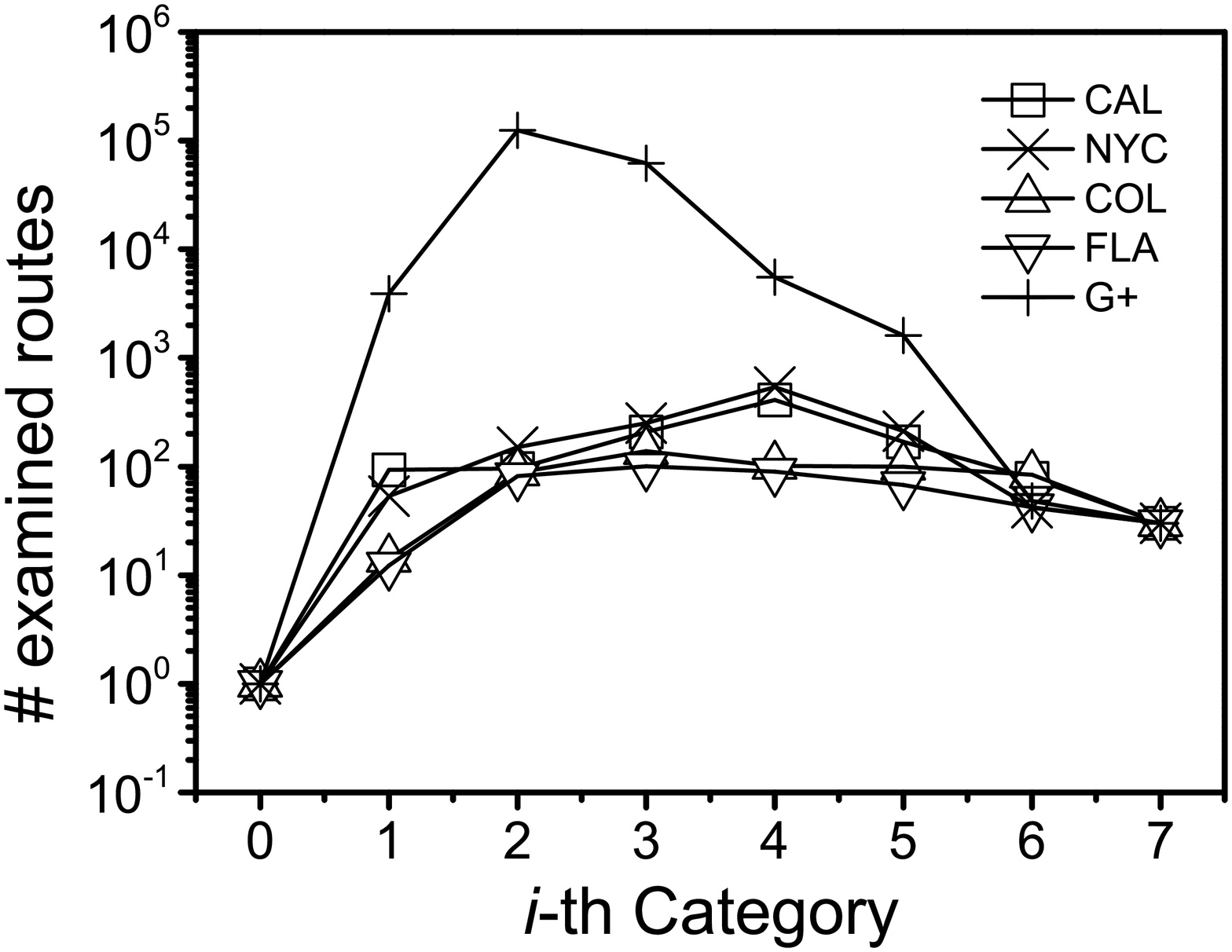}
\caption{Searching space of SK at different categories}
\label{space}
\end{minipage}%
\hspace{0.1in}
\begin{minipage}[t]{0.45\linewidth}
\centering
\includegraphics[width=1.6in]{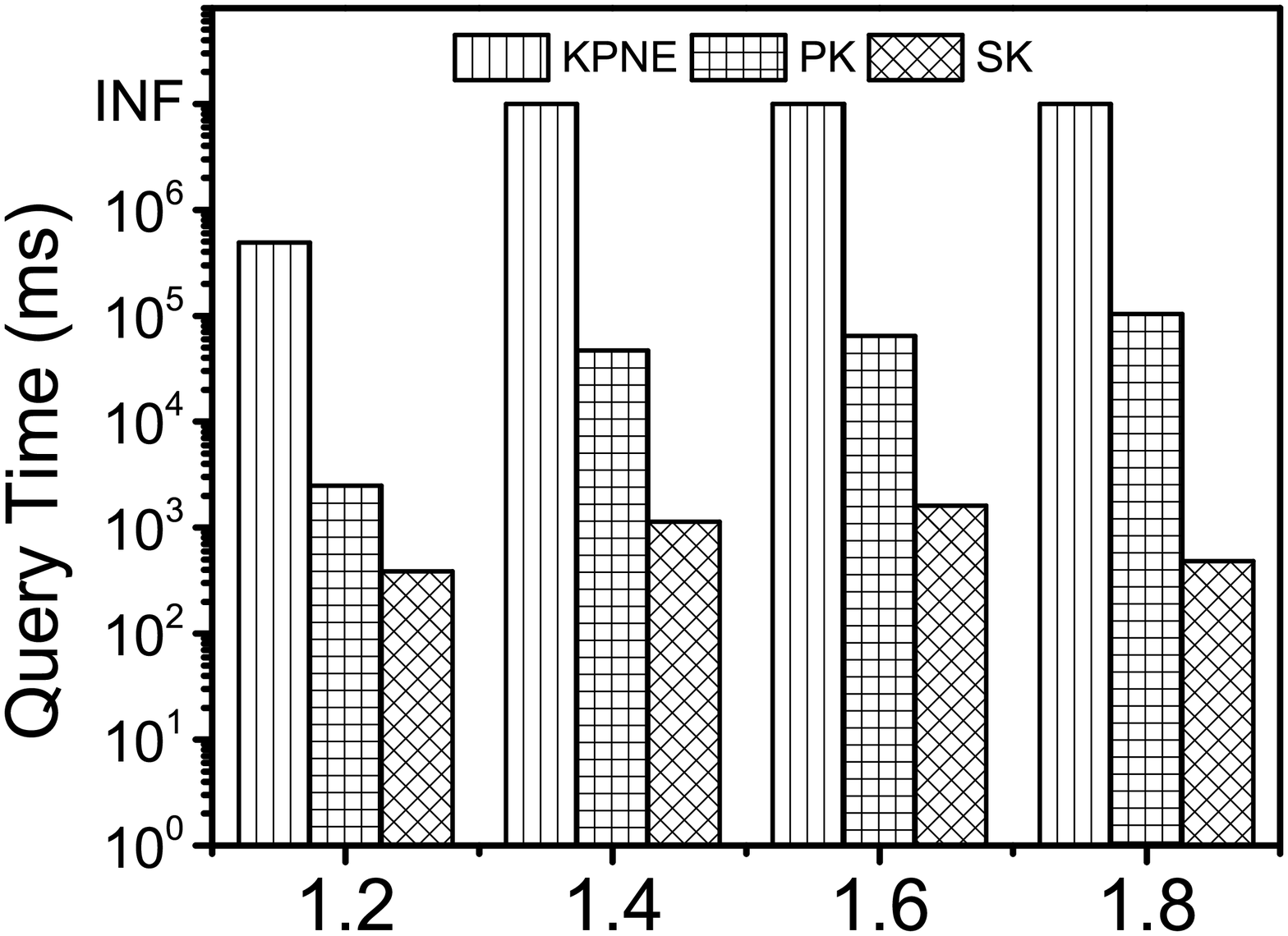}
\caption{Performance on zipfian distribution}
\label{zip}
\end{minipage}
\vspace*{-0.22in}
\end{figure}

Figure \ref{space} shows the searching space of SK at different categories on different graphs. Initially, only one route (source $s$) is examined at category 0. Then the number of examined routes increases along the category sequence, because the estimated costs are loose and more candidate routes are enabled to be examined. As the estimated costs are closer and closer to the real least costs, the number of examined routes quickly decreases and the searching space shrinks. Finally, only $30$ routes are examined at the last category (i.e., for destination $t$). The searching space begins to decrease at the 3rd or 4th category or even earlier, which is very efficient.
Figure \ref{space} is also consistent with the intuition shown in Figure~\ref{spaceF}(c).

Next, we show performance while varying important parameters. Due to the space limitation, we only report experimental results on small graph $\mathit{CAL}$ with real categories and large graph $\mathit{FLA}$ with synthetic categories.

\begin{figure}[t]
\centering
\subfigure[$\mathit{CAL}$]{
\label{divTime}
\includegraphics[width=1.6in]{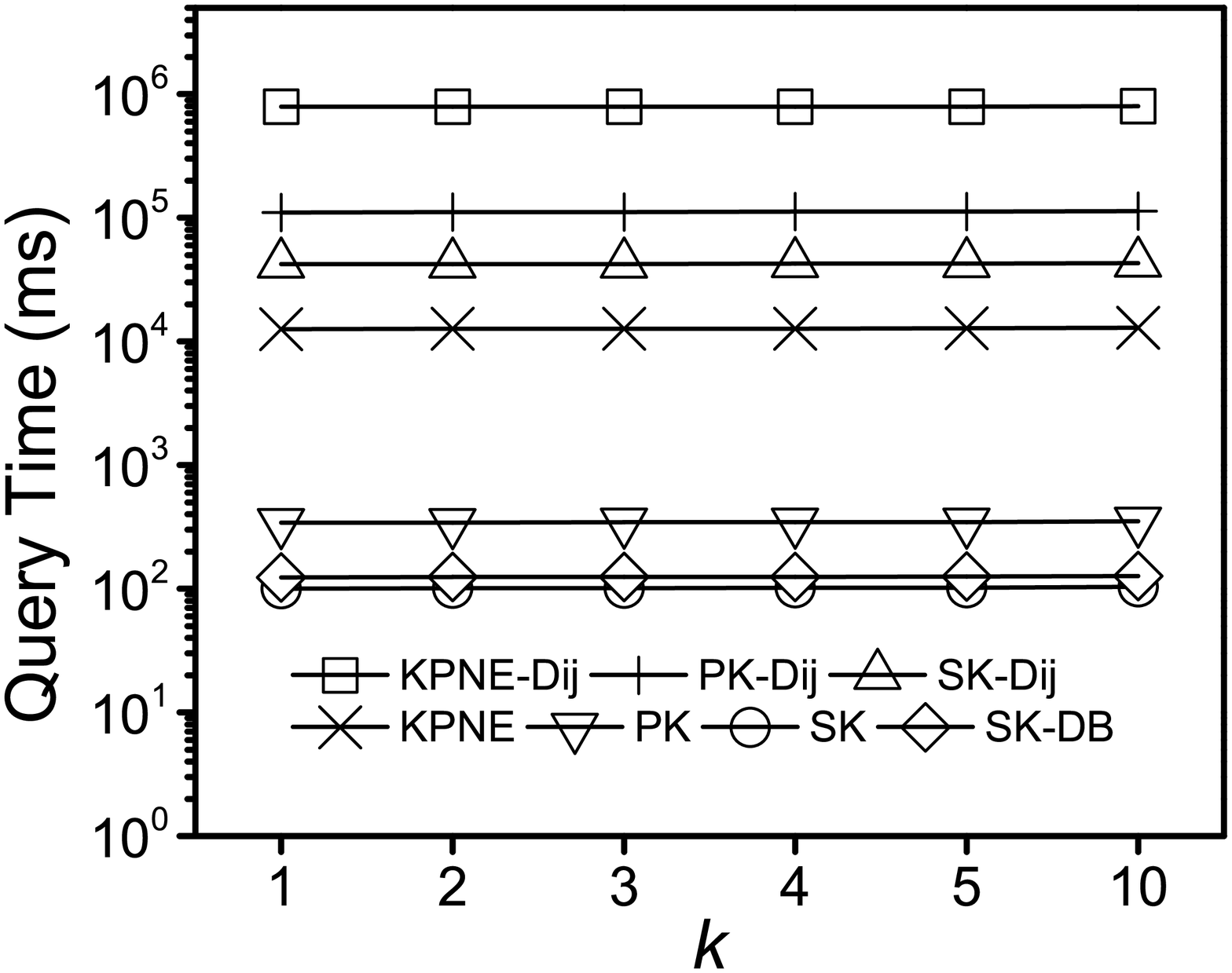}}
\subfigure[$\mathit{FLA}$]{
\label{divPath}
\includegraphics[width=1.6in]{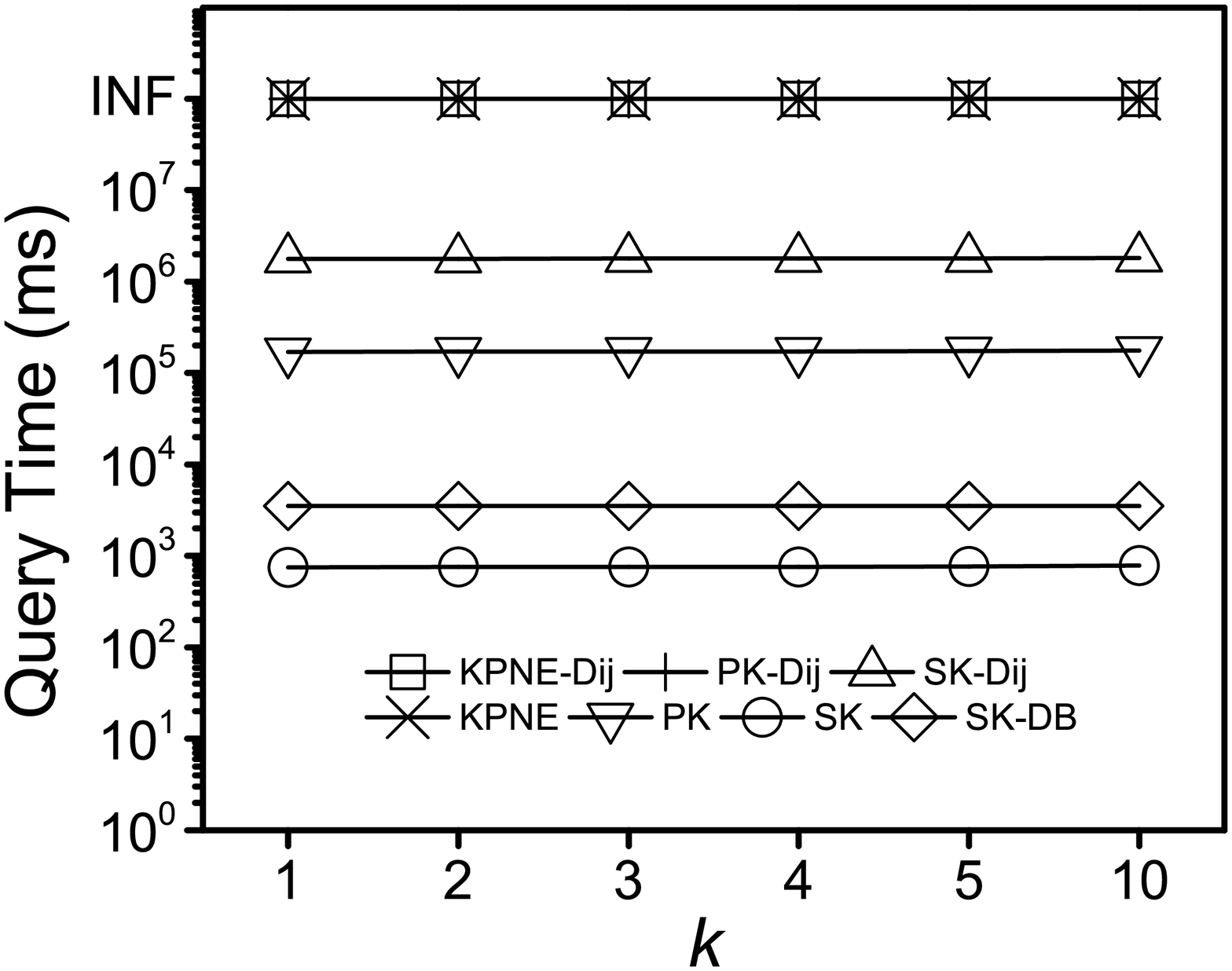}}
\caption{The performance of different methods with small $k$}
\label{smallk}
\vspace*{-0.2in}
\end{figure}

\noindent
\textbf{Effect of $k$.} Figures \ref{flak} and \ref{calk} show the effect of parameter $k$. On large graph $\mathit{FLA}$, KPNE, KPNE-Dij, and PK-Dij cannot return the results within 3,600 seconds, even when $k=10$, due to larger searching space or too many NN queries by Dijkstra's search. On small graph $\mathit{CAL}$, all methods are able to compute the results and KPNE, PK, SK are much more efficient than KPNE-Dij, PK-Dij, SK-Dij, respectively. Both Figures \ref{flak} and \ref{calk} show that SK and SK-DB greatly outperform other methods in different $k$ due to much fewer examined routes and NN queries. Note that all methods perform steadily with different $k$, meaning that they are scalable w.r.t. $k$ and are able to process KOSR with large $k$s. This is because the top-$k$ optimal sequenced routes tend to have similar costs, and once we find the 1st optimal sequenced route, other optimal sequenced routes are also considerably covered in its searching space. As a result, fewer NN queries are needed to find the other optimal sequenced routes when $k$ rises, therefore, the query time only increases slightly, which is also consistent with the time complexity analysis in Lemma \ref{complexity}. Figure \ref{smallk} shows the performance of different methods with small $k$ on $\mathit{CAL}$ and $\mathit{FLA}$. The query time of all methods slightly change as $k$ slowly increases, and our proposed algorithms also outperform existing algorithms.

\noindent
\textbf{Effect of $|C|$.} The performance of different methods on $\mathit{FLA}$ and $\mathit{CAL}$ by varying the size of category sequence $|C|$ is shown in Figures \ref{flacat} and \ref{calcat}. When $|C|=2$, KPNE is able to return the results on $\mathit{FLA}$ as the searching space is small. However, KPNE-Dij and PK-Dij still cannot compute the results on $\mathit{FLA}$ within 3,600 seconds due to too many Dijkstra's searches on large graphs for KPNE and PK. As $|C|$ increases, the searching space of KPNE increases exponentially, KPNE fails to return the results on $\mathit{FLA}$ when $|C|\ge 4$. On small graph $\mathit{CAL}$, KPNE-Dij cannot return results when $|C|\ge 8$. Although the searching spaces and run-times of PK and SK (SK-DB) increase as $|C|$ gets larger due to greater $M$ and $N$ in Lemma \ref{complexity}, SK (SK-DB) greatly outperforms PK in all settings. In addition, the run-time of SK (SK-DB) increases more slowly than PK.
A larger $|C|$ means more label indexes need to be loaded into memory and initialized by SK-DB. As a result, SK-DB needs more disk accesses and thus a higher overhead compared to SK.

\noindent
\textbf{Effect of $|C_i|$.} Figure \ref{flacap} shows the performance of different methods on $\mathit{FLA}$ by varying the size of vertices in each category, i.e., $|C_i|$. We only report experiments on the largest graph $\mathit{FLA}$ as we do not generate categories for $\mathit{CAL}$. Due to the huge searching space, KPNE, KPNE-Dij, and PK-Dij cannot return the results even when $|C_i|=5,000$. Obviously, the performance of both PK and SK deteriorates as $|C_i|$ increases, because the time complexity of the two methods increases as $|C_i|$ increases according to Lemma \ref{complexity}. Intuitively, a larger $|C_i|$ means more vertices in each category and thus more routes to be examined. Clearly, SK is more efficient than PK due to much fewer NN queries. As $|C_i|$ increases, the runtime increasing trend of SK(-DB) is slower than that of PK, which means SK(-DB) is more scalable w.r.t. $|C_i|$.

\noindent
\textbf{Zipfian category distribution.} Figure \ref{zip} illustrates the results of different $f$ with $|C|=6, k=30$ on $\mathit{FLA}$. Clearly, our methods greatly outperforms baseline KPNE in all settings. It shows that the query time of PK increases as $f$ gets larger, and KPNE cannot return the results when $f\ge 1.4$. This is because a larger $f$ means less skew distribution. As a result, the number of partially explored routes to be examined between consecutive categories, i.e., $|C_i|\cdot |C_{i+1}|$ in the worst case (see Lemma \ref{complexity}), gets larger for less skew distribution since $|C_i|$ and $|C_i+1|$ tend to be similar in this case. Hence, more time is needed to find the top-$k$ optimal sequenced routes. Moreover, since SK filters much more routes, it greatly outperforms PK.



\begin{wrapfigure}{r}[0cm]{0pt}
\includegraphics[width=1.6in]{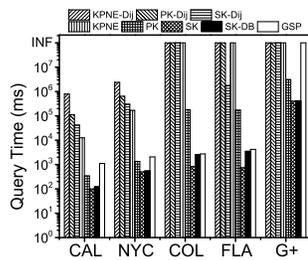}
\caption{Performance of different methods for OSR queries}
\label{ssr}
\vspace*{-0.22in}
\end{wrapfigure}

\noindent
\textbf{Performance for the OSR queries.} By setting $k=1$, the KOSR problem becomes the OSR problem. We evaluate the performance of the state-of-the-art OSR method, GSP, and our proposed methods. Figure \ref{ssr} shows the run-time of different methods on different graphs. Clearly, the state-of-the-art method GSP outperforms KPNE(-Dij), PK-Dij and SK-Dij on all graphs, and GSP also outperforms PK on graphs with large category size, i.e., $\mathit{COL}$, and $\mathit{FLA}$, because GSP only requires $O(|C|)$ graph searches to find the optimal sequenced route, while PK needs much more examined routes and NN queries on these graphs. However, on graph with small category size, i.e., $\mathit{CAL}$ and $\mathit{NYC}$, PK is more efficient than GSP due to fewer examined routes and NN queries. In all settings, SK and SK-DB are more efficient than GSP, since SK and SK-DB have much smaller searching space by using the target-directed cost estimation strategy, and achieve very efficient NN query by using inverted label indexes. 
For $\mathit{G+}$, we cannot build the contraction hierarchy structure for GSP on $\mathit{G+}$ in 3 days, thus GSP cannot return the results on $\mathit{G+}$. In addition, the run-time of GSP is dependent on the graph sizes. As the graph size increases, GSP takes longer time. In contrast, the runtime of SK(-DB) is independent of the graph sizes, meaning that it has better scalability w.r.t. the graph sizes. 

\section{Conclusion and Outlook} \label{sec:con}
In this paper, we study the top-$k$ optimal sequenced routes problem. We propose efficient algorithms based on a novel route dominance relationship and a target-directed cost estimation strategy using hop labeling techniques. Extensive experiments on real world graphs demonstrate that the proposed algorithms are efficient.

  As a future work, we plan to fill the gaps as shown in Table~\ref{rw} to solve the KOSR querying when partial or arbitrary category orders or personal preferences for categories are allowed on both Euclidean and general graphs.

\section*{Acknowledgement}
Our research is supported by the National Key Research and Development Program of China (2016YFB1000905), NSFC (61532021, U1501252, 61702423, and 61772327).


\end{document}